\documentclass[10pt, journal]{IEEEtran}
\usepackage{amsmath,amsfonts}
\usepackage{array}
\usepackage{textcomp}
\usepackage{stfloats}
\usepackage{url}
\usepackage{verbatim}
\usepackage{hyperref}
\usepackage{multirow}
\usepackage{float}
\usepackage[caption = false]{subfig}
\usepackage[final]{graphicx}
\usepackage[ruled,linesnumbered]{algorithm2e}
\def\BibTeX{{\rm B\kern-.05em{\sc i\kern-.025em b}\kern-.08em
    T\kern-.1667em\lower.7ex\hbox{E}\kern-.125emX}}
\usepackage{balance}
\hypersetup{draft}

\begin{document}
\title{White-Box 3D-OMP-Transformer for ISAC}
\author{Bowen Zhang and Geoffrey Ye Li\\
\small Department of Electrical and Electronic Engineering, Imperial College London, London, UK\\
\small \{k.zhang21, geoffrey.li\}@imperial.ac.uk
}
\maketitle

\begin{abstract}
Transformers have found broad applications for their great ability to capture long-range dependency among the inputs using attention mechanisms. The recent success of transformers increases the need for mathematical interpretation of their underlying working mechanisms, leading to the development of a family of white-box transformer-like deep network architectures. However, designing white-box transformers with efficient three-dimensional (3D) attention is still an open challenge. In this work, we revisit the 3D-orthogonal matching pursuit (OMP) algorithm and demonstrate that the operation of 3D-OMP is analogous to a specific kind of transformer with 3D attention. Therefore, we build a white-box 3D-OMP-transformer by introducing additional learnable parameters to 3D-OMP. As a transformer, its 3D-attention can be mathematically interpreted from 3D-OMP; while as a variant of OMP, it can learn to improve the matching pursuit process from data. Besides, a transformer's performance can be improved by stacking more transformer blocks. To simulate this process, we design a cascaded 3D-OMP-Transformer with dynamic small-scale dictionaries, which can improve the performance of the 3D-OMP-Transformer with low costs. We evaluate the designed 3D-OMP-transformer in the multi-target detection task of integrated sensing and communications (ISAC). Experimental results show that the designed 3D-OMP-Transformer can outperform current baselines.

\end{abstract}
%
%
\section{Introduction}
\label{sec:intro} 
Transformer and its variants have been used extensively in natural language processing (NLP) and computer vision (CV) tasks \cite{devlin2018bert, dosovitskiy2020image, liu2021swin} since proposed in \cite{vaswani2017attention}. Transformers are incredibly successful because of their attention mechanisms, which can learn layer-wise queries, keys, and values to represent long-range correlation in the data. Current attention techniques primarily capture the dependency in a particular dimension, such as the sequential dimension of tokens in vanilla transformers \cite{vaswani2017attention}, the spatial dimension of images in visual transformers \cite{dosovitskiy2020image, liu2021swin}, and the spectral dimension of hyperspectral images in spectral-wise transformers \cite{cai2022mask}. When capturing the dependency in three-dimensional (3D) data, current works stack multiple types of transformers sequentially, including spectral and spatial transformers in \cite{zhong2021spectral}. Since 3D correlation is not thoroughly investigated, it is unknown whether this technique is efficient in modelling 3D data. Therefore, designing transformers with efficient 3D attention is still an open question.

Recent advances in transformers also increase the need for mathematical interpretation of transformers' working mechanisms \cite{englebert2023explaining}. Several works on white-box transformers have been conducted to satisfy this demand \cite{yu2024white, de2023designing, yang2022transformers}. These works interpret transformers as an unfolded optimization process. For example, Yang et. al. \cite{yang2022transformers} have explained the transformers' forward pass as the optimization process of a particular energy function via alternating inexact minimization (AIM). Yu et al. \cite{yu2024white} have interpreted transformers as the optimization of a sparse rate reduction objective function. In their approach, the self-attention layer minimizes the lossy coding rate while the feed-forward network (FFN) induces sparsity. These interpretations advance our understanding of transformers and inspire new transformer designs \cite{de2023designing}. 

In this work, we revisit the 3D-orthogonal matching pursuit (3D-OMP) algorithm and demonstrate that the operation of 3D-OMP can is analogous to a transformer with external attention mechanisms \cite{guo2022beyond} and 3D attention. We then build a 3D-OMP-Transformer by introducing additional learnable parameters into 3D-OMP. As a variant of 3D-OMP, the matching pursuit process can be improved through learning from data. As a transformer, the proposed 3D-OMP-Transformer can efficiently capture 3D dependency inside the inputs by three sets of domain-specific dictionaries. Furthermore, considering the fact that a transformer's performance can be improved by stacking more transformer blocks, we design a cascaded 3D-OMP-Transformer with dynamic small-scale dictionaries. In our formulation, the attention mechanisms of the transformers can be viewed as finding the atom that maximally reduces the approximation error and the FFN can be viewed as trying to subtract the influence of the atom from the original signal. Stacking Transformer blocks are interpreted as refining dictionaries of 3D-OMP. 

The proposed white-box 3D-OMP-Transformer is designed for the multi-target detection
task of integrated sensing and communications (ISAC) \cite{liu2020joint,zhang2018multibeam,mateos2023model}. ISAC requires to estimate the statuses of targets from the echo signal of a communication waveform in the 3D (angle-delay-Doppler) domain \cite{liu2020joint}. As will be explained later in Eq. (\ref{equ:6}), the influence of one target on the echo signal is distributed over the whole 3D domain; therefore, capturing 3D long-range dependency using attention can improve the detection accuracy in ISAC. Since the influence can be well-modelled, exploring white-box networks allows us to utilize the domain knowledge from the model \cite{zhang2018multibeam,mateos2023model} and reduces the time of network training. 

In addition to our work, there have been some attempts to use neural networks for the multi-target detection task in ISAC \cite{mateos2023model, muth2023autoencoder,mateos2022end}. These works have demonstrated the benefits of using neural networks to calibrate and compensate for hardware imperfection in localization measurements, including antenna misplacement and clock offsets. Different from existing works, we show that the neural network-based solution can outperform current baselines even when there is no imperfection. 

Note that a differentiable OMP has been proposed in \cite{mateos2023model}. By introducing a softmax operation, Mateos-Ramo et. al. \cite{mateos2023model} have successfully unfolded OMP and made it differentiable. Different from \cite{mateos2023model}, we incorporate OMP into transformers and achieve better performance. 

\section{ISAC system settings and model}
In this section, we will introduce the system settings and models of the considered ISAC scenario.
\subsection{System Settings}
Suppose that an ISAC transmitter is trying to localize $M$ passive targets in an environment using a reflected communication waveform. The angle-of-arrival (AoA), ranges, and velocities of the targets are uniformly distributed in $[\phi_{\text{min}}, \phi_{\text{max}}]$, $[r_{\text{min}}, r_{\text{max}}]$, and $[v_{\text{min}}, v_{\text{max}}]$, respectively. We assume that the ISAC transmitter works in multiple-input-multiple-output (MIMO) configuration with a uniform linear array (ULA) of $K$ antennas with an adjacent antenna distance $d$. We assume that the ISAC transmitter uses OFDM with $S$ sub-carriers and $T_{s}$ OFDM symbols in a time slot. For OFDM modulation, the sub-carrier space is $\Delta_{f}$ and the symbol duration is $\Delta_{T}$. The carrier frequency is denoted as $f_{c}$ and wavelength is $\lambda=c/f_{c}$, where $c$ is the speed of light. We further assume that $d=\lambda/2$. The considered ISAC scenario is similar to the one in \cite{mateos2023model}. 


\subsection{System Models}
Assume the ISAC transmitter is sending a sequence of complex modulated symbols $\mathbf{Y}\in \mathcal{\textit{C}}^{S\times T_{s}}$ to the communication user using 4-quadrature amplitude modulation (4-QAM). 
Denote $Y_{ij}$ as the symbol transmitted at the $i$-th sub-carrier of the $j$-th OFDM symbols. The signal is transmitted using $K$ antennas in MIMO systems. To improve the sensing performance, the ISAC transmitter uses a transmitting beamformer, $\mathbf{p}\in \mathcal{C}^{K\times 1}$, based on the AoA ranges of targets. Specifically, $\mathbf{p}$ is the solution of $\text{min}_{\mathbf{y}}||\mathbf{b}-\mathbf{A^{T}}\mathbf{y}||$, where $\mathbf{b}\in \mathcal{R}^{N_{\phi\text{grid}}\times 1}$ denotes the desired
beampattern at $N_{\phi\text{grid}}$ angular grid locations $
\{\phi_{i}\}_{i=1}^{N_{\phi\text{grid}}}$ and satisfies,

\begin{equation}
b_{i}=\begin{cases}
     1 ,& \text{if } \phi_{i}\in [\phi_{\text{min}}, \phi_{\text{max}}],\\
    0,              & \text{otherwise},
\end{cases}
\label{equ:add}
\end{equation}
$\mathbf{A}=[\boldsymbol{a}_{tx}(\phi_{1}),\boldsymbol{a}_{tx}(\phi_{2}),...,\boldsymbol{a}_{tx}(\phi_{N_{\phi\text{grid}}})]\in \mathcal{R}^{K \times N_{\phi\text{grid}}}$ consists of the transmitting steering vectors,  
\begin{equation}
\begin{aligned}
\boldsymbol{a}_{tx}(\phi)=[&e^{-j2\pi d\sin{\phi}/\lambda}, e^{-j2\pi 2d\sin{\phi}/\lambda}, \dots, \\&
e^{-j2\pi Kd\sin{\phi}/\lambda}]^T.
\end{aligned}
\label{equ:2}
\end{equation}
The reflected signal, $\mathbf{Z}$, from $M$ targets can be expressed as, 
\begin{equation}
\begin{aligned}
\mathbf{Z} &=\sum_{m=1}^{M}\alpha_{m}\boldsymbol{a}_{rx}(\phi_{m})\otimes \{(\boldsymbol{a}_{tx}^{T}(\phi_{m})\mathbf{p})[\mathbf{Y}\odot \\&(\boldsymbol{\rho}(r_{m})\boldsymbol{\beta}(v_{m})^{T})]\}+\boldsymbol{W}.
\end{aligned}
\label{equ:3}
\end{equation}
where $\otimes$ and $\odot$ denote the outer product and the element-wise multiplication, respectively. $T$ denotes the transpose of the vector. $\mathbf{Z}\in \mathcal{C}^{K\times S \times T_{s}}$ is the received signal in the spatial-frequency-time domain. $\alpha_{m} \sim \mathcal{CN}(0, \sigma_{s}^{2})$ is the complex channel gain of the $m$-th target. $\boldsymbol{W}\in \mathcal{C}^{K\times S \times T_{s}}$ is the channel noise with $W_{ijk}\sim \mathcal{CN}(0, N_{0})$, where $N_{0}$ determines the sensing signal-to-noise ratio (SNR) by $\text{SNR}_{s}=10\log_{10}(K\sigma_{s}^{2}/N_{0})$ dB, following \cite{mateos2023model}. $\boldsymbol{a}_{rx}(\phi_{m})\in \mathcal{C}^{K\times 1}$ and $\boldsymbol{a}_{tx}(\phi_{m})\in \mathcal{C}^{K\times 1}$ are the receiving and transmitting steering vectors of target $m$, respectively. Their expressions are given in Eq. (\ref{equ:2}). Denote $r_{m}$ and $v_{m}$ as the range and velocity of target $m$, respectively. The impact of range, $r_{m}$, in different sub-carriers is reflected by $\boldsymbol{\rho}(r_{m})\in \mathcal{C}^{S\times 1}$ through the delay $\tau_{m}=\frac{2r_{m}}{c}$ as,
\begin{equation}
\boldsymbol{\rho}(r_{m})=[e^{-j2\pi \Delta_{f}\tau_{m}}, e^{-j2\pi 2\Delta_{f}\tau_{m}}, \cdots, e^{-j2\pi S\Delta_{f}\tau_{m}}]^T.
\label{equ:4}
\end{equation}
Velocity, $v_{m}$, will affect different OFDM symbols by $\boldsymbol{\beta}(v_{m})\in \mathcal{C}^{T_{s}\times 1}$ through the Doppler frequency $f_{dm}=\frac{2v_{m}f_{c}}{c}$ as,
\begin{equation}
\boldsymbol{\beta}(v_{m})=[e^{j2\pi \Delta_{T}f_{dm}}, e^{j2\pi 2\Delta_{T}f_{dm}}, \cdots, e^{j2\pi T_{s}\Delta_{T}f_{dm}}]^T.
\label{equ:5}
\end{equation}


After receiving the echo signal, the impact of the transmitted signal, $\boldsymbol{Y}$, is first removed by element-wise division as,
\begin{equation}
\boldsymbol{\hat{Z}}=\boldsymbol{Z}\oslash (\mathbf{1}\otimes \boldsymbol{Y})=\sum_{m=1}^{M} \gamma_{m} \boldsymbol{a}_{rx}(\phi_{m}) \otimes \boldsymbol{\rho}(r_{m})  \otimes \boldsymbol{\beta}(v_{m})+\boldsymbol{\hat{W}}.
\label{equ:6}
\end{equation}
where $\oslash$ denotes the element-wise division, $\mathbf{1} \in \mathcal{C}^{K\times 1}$, $\gamma_{m}=\alpha_{m}(\boldsymbol{a}_{tx}^{T}(\phi_{m})\boldsymbol{p})$, $\boldsymbol{\hat{W}}=\boldsymbol{W}\oslash (\mathbf{1}\otimes \boldsymbol{Y})$. $\boldsymbol{\hat{Z}}\in \mathcal{C}^{K\times S \times T_{s}}$ is then passed to the multi-target detection algorithm. Basically, the algorithm aims to estimate $\{\phi_{m}, r_{m}, v_{m}\}_{m=1}^{M}$ from $\boldsymbol{\hat{Z}}$.


\section{3D-OMP: A Special Case of Transformer}
In this section, we introduce the 3D-OMP algorithm for multi-target detection and describe the basics of transformers. Then, we show that the operation of the 3D-OMP algorithm is analogous to a specific kind of transformer. 

\subsection{3D-OMP for ISAC}
The 3D parameters of the targets can be obtained by the 3D-OMP algorithm \cite{mateos2023model}. To construct an over-complete dictionary for OMP, we first specify an angle grid $\boldsymbol{G}_{\phi}=[\phi_{1}, \phi_{2}, \cdots, \phi_{N_{\phi}}] \in \mathcal{R}^{N_{\phi}\times 1}$, a range grid $\boldsymbol{G}_{r}=[r_{1}, r_{2}, \cdots, r_{N_{r}}] \in \mathcal{R}^{N_{r}\times 1}$, and a velocity grid $\boldsymbol{G}_{v}=[v_{1}, v_{2}, \cdots, v_{N_{v}}] \in \mathcal{R}^{N_{v}\times 1}$. The angles, ranges, velocities inside $\boldsymbol{G}_{\phi}$, $\boldsymbol{G}_{r}$, $\boldsymbol{G}_{v}$, distribute uniformly over $[\phi_{min}, \phi_{max}]$, $[r_{min}, r_{max}]$, and $[v_{min}, v_{max}]$, respectively. Then, the angle, range, and velocity dictionaries can be defined as,

\begin{equation}
\begin{split}
\boldsymbol{\Phi}_{\phi} &=[\boldsymbol{a}(\phi_{1}), \cdots, \boldsymbol{a}(\phi_{N_{\phi}})] \in \mathcal{C}^{K\times N_{\phi}} \\
\boldsymbol{\Phi}_{r} &=[\boldsymbol{\rho}(r_{1}), \cdots, \boldsymbol{\rho}(r_{N_{r}})] \in \mathcal{C}^{S \times N_{r}}
\\
\boldsymbol{\Phi}_{v} &=[\boldsymbol{\beta}(v_{1}), \cdots, \boldsymbol{\beta}(v_{N_{v}})] \in \mathcal{C}^{T_{s} \times N_{v}}
\end{split}
\label{equ:7}
\end{equation}
With the dictionaries in Eq. (\ref{equ:7}), Eq. (\ref{equ:6}) can be rewritten as,
\begin{equation}
\boldsymbol{\tilde{Z}}=\boldsymbol{D}\boldsymbol{s}+\boldsymbol{\hat{W}},
\label{equ:8}
\end{equation}
where $\boldsymbol{\tilde{Z}} \in \mathcal{C}^{KST_{s}\times 1}$ is the vector form of $\boldsymbol{\hat{Z}}$, $\boldsymbol{D} = \boldsymbol{\Phi}_{\phi}\otimes \boldsymbol{\Phi}_{r} \otimes \boldsymbol{\Phi}_{v} \in \mathcal{C}^{KST_{s} \times N_{\phi}N_{r}N_{v}}$ is the integrated dictionary, $\boldsymbol{s} \in \mathcal{C}^{N_{\phi}N_{r}N_{v}\times 1}$ is the sparse vector whose non-zero elements' positions represent the desired parameters. As the size of the dictionary, $D$, is extremely large in ISAC, 3D-OMP is more suitable to solve Eq. (\ref{equ:8}). The 3D-OMP algorithm is summarized in Algorithm \ref{alg:one}. Operator $\hat{\boldsymbol{Z}}\Delta \boldsymbol{\Phi}_{\phi}^{*}$ in line 1 calculates the tensor product of $\hat{\boldsymbol{Z}}$ and the complex conjugate of $\boldsymbol{\Phi}_{\phi}$, contracting the first dimensions of $\hat{Z}$ and $\boldsymbol{\Phi}_{\phi}^{*}$. The resulting tensor contains the uncontracted dimensions of $\hat{\boldsymbol{Z}}$ followed by the uncontracted dimensions of $\boldsymbol{\Phi}_{\phi}^{*}$. This process can be depicted as,
\begin{equation}
[\hat{\boldsymbol{Z}}\Delta \boldsymbol{\Phi}_{\phi}^{*}]_{i,j,k}=\sum_{m=1}^{K}[\hat{\boldsymbol{Z}}]_{m,i,j}[\boldsymbol{\Phi}_{\phi}^{*}]_{m,k}.
\label{equ:9}
\end{equation}
Operator $|\cdot|$ in line 1 calculates the absolute value of the complex numbers in a tensor.

At each iteration, the 3D-OMP searches for the atom, $\boldsymbol{\varphi}_{m}=\boldsymbol{a}_{rx}(\phi_{i}) \otimes \boldsymbol{\rho}(r_{j})  \otimes \boldsymbol{\beta}(v_{k})$, that is most correlated with the residual $\hat{\boldsymbol{Z}}^{(m)}$ by line 3, keeps track of the active atom set, $\mathcal{M}_{m+1}=\mathcal{M}_{m} \cup \boldsymbol{\varphi}_{m}$, by line 5, computes the projection of the signal, $\hat{\boldsymbol{Z}}$, onto $\mathcal{M}_{m+1}$ by line 6, and uses the residual for the new iteration by line 7.

\SetKwInput{KwInput}{Input}                
\SetKwInput{KwOutput}{Output}
\SetKwInput{KwInitial}{Initialization}
\SetKwComment{Comment}{/* }{ */}
\begin{algorithm}[t!]
\caption{3D-OMP for Multi-target Detection}\label{alg:one}

\KwInput{Observation $\boldsymbol{\hat{Z}}$ in (\ref{equ:6}), Dictionaries $\boldsymbol{\Phi}_{\phi}$, $\boldsymbol{\Phi}_{r}$, $\boldsymbol{\Phi}_{v}$, grids $\boldsymbol{G}_{\phi}$, $\boldsymbol{G}_{r}$, $\boldsymbol{G}_{v}$, termination threshold $\delta$.}
\KwOutput{Set $\mathcal{T}=\{\phi_{m}, r_{m}, v_{m}\}_{m=1}^{M}$}
\KwInitial{$m=1$, $\boldsymbol{\hat{Z}}^{(1)}=\boldsymbol{\hat{Z}}$, $\mathcal{T}=\emptyset$, $\mathcal{M}=\emptyset$}
Compute the angle-delay-Doppler map $\mathcal{L}(\boldsymbol{\hat{Z}})=|\boldsymbol{\hat{Z}} \Delta \boldsymbol{\Phi}_{\phi}^{*} \Delta \boldsymbol{\Phi}_{r}^{*} \Delta \boldsymbol{\Phi}_{v}^{*}|\in \mathcal{R}^{N_{\phi}\times N_{r} \times N_{v}}$;

\While{$\text{max} \, \mathcal{L}(\boldsymbol{\hat{Z}}^{(m)})>\delta$}{
  Find $(i,j,k)=\text{argmax}_{i,j,k}\mathcal{L}(\boldsymbol{\hat{Z}}^{(m)})$ 
  
  Update target set $\mathcal{T}\leftarrow \mathcal{T} \cup \{\phi_{m}=[\boldsymbol{G}_{\phi}]_{i}, r_{m}=[\boldsymbol{G}_{r}]_{j}, v_{m}=[\boldsymbol{G}_{v}]_{k}\}$   
  
  Update atom set $\mathcal{M} \leftarrow \mathcal{M} \cup \boldsymbol{\varphi}_{m}=\{\boldsymbol{a}_{rx}(\phi_{i}) \otimes \boldsymbol{\rho}(r_{j})  \otimes \boldsymbol{\beta}(v_{k})\}$

  Update gain estimate $\boldsymbol{\hat{\gamma}}=\text{argmin}_{\boldsymbol{\gamma}} \|\boldsymbol{\hat{Z}}-\sum_{i=1}^{m} \gamma_{i} \boldsymbol{\varphi}_{i}\|$

  Update residual
  $\boldsymbol{\hat{Z}}^{(m+1)}=\boldsymbol{\hat{Z}}-\sum_{i=1}^{m} \hat{\gamma}_{i} \boldsymbol{\varphi}_{i}$

  $m=m+1$
  
}

\end{algorithm}


\subsection{Transformer}
A transformer layer is typically composed of an attention layer and a feed-forward network (FFN) layer. The attention layer is built on either self-attention or external attention in different transformers \cite{yang2022transformers, guo2022beyond}. We first revisit the transformer with self-attention \cite{vaswani2017attention}. Given an input feature map, $\boldsymbol{F} \in \mathcal{R}^{N\times d}$, where $N$ is the number of elements and $d$ is the number of features for each element, self-attention linearly projects $\boldsymbol{F}$ to a query matrix, $\boldsymbol{Q} \in \mathcal{R}^{N\times d'}$, a key matrix, $\boldsymbol{K} \in \mathcal{R}^{N\times d'}$, and a value matrix, $\boldsymbol{V} \in \mathcal{R}^{N\times d}$, by $\boldsymbol{Q}=\boldsymbol{F}\boldsymbol{W_{Q}}$, $\boldsymbol{K}=\boldsymbol{F}\boldsymbol{W_{K}}$, and $\boldsymbol{V}=\boldsymbol{F}\boldsymbol{W_{V}}$. Then the self-attention layer can be formulated as,
\begin{equation}
\boldsymbol{A}=\text{softmax}(\sigma \boldsymbol{Q}\boldsymbol{K}^{T}),
\boldsymbol{F}_{attn}=\boldsymbol{A}\boldsymbol{V},
\label{equ:10}
\end{equation}
where $\boldsymbol{A} \in \mathcal{R}^{N\times N}$ is the attention map, $\sigma$ is the reweighting coefficient of the softmax operator, $\boldsymbol{F}_{attn} \in \mathcal{R}^{N\times d}$ is the output feature map of the self-attention layer. After the self-attention layer, $\boldsymbol{F}_{attn}$ is fed to an FFN layer,
\begin{equation}
\boldsymbol{F}_{out}=\text{FFN}(\boldsymbol{F}_{attn}),
\label{equ:11}
\end{equation}
where $\boldsymbol{F}_{out}$ is the output of a transformer layer. An FFN layer consists of linear transformations and non-linear activations.

For the transformers with external attention \cite{guo2022beyond}, $\boldsymbol{K}$ and $\boldsymbol{V}$ are not generated from $\boldsymbol{F}$ by linear projection. Instead, they are independent of input features and shared across the entire dataset, called external memory units. The external attention can lead to lightweight architectures when the number of
elements in the external memories is much smaller than that in the input features.

\subsection{3D-OMP Represented in the Form of Transformers}
Although the 3D-OMP algorithm and transformers are designed for different signal recovery problems, the operation of 3D-OMP is analogous to transformers with external attention. To explain this, we rewrite the 3D-OMP algorithm in the formulation of the attention layer (represented by Eq. (\ref{equ:10})) and the FFN layer (represented by Eq. (\ref{equ:11})). 

To construct the attention layer, we first treat the residual at each iteration as queries, i.e., $\boldsymbol{Q}=\boldsymbol{\hat{Z}}^{(m)}$. We then define keys as the Hermitian matrices of dictionaries, i.e. $\boldsymbol{K}_{\phi}=\boldsymbol{\Phi}_{\phi}^{H}$, $\boldsymbol{K}_{r}=\boldsymbol{\Phi}_{r}^{H}$, and $\boldsymbol{K}_{v}=\boldsymbol{\Phi}_{v}^{H}$. Next, we define two different sets of values. One set is from grids as $\boldsymbol{V}_{\phi}^{(1)}=\boldsymbol{G}_{\phi}$, $\boldsymbol{V}_{r}^{(1)}=\boldsymbol{G}_{r}$, and $\boldsymbol{V}_{v}^{(1)}=\boldsymbol{G}_{v}$, another is from dictionaries as $\boldsymbol{V}_{\phi}^{(2)}=\boldsymbol{\Phi}_{\phi}$, $\boldsymbol{V}_{r}^{(2)}=\boldsymbol{\Phi}_{r}$, and $\boldsymbol{V}_{v}^{(2)}=\boldsymbol{\Phi}_{v}$. With these definitions, line 3 in Algorithm \ref{alg:one} can be modelled as the process of calculating a 3D attention map from queries and keys by, 
\begin{equation}
\begin{split}
\boldsymbol{A}&=\text{softmax}(\sigma \boldsymbol{Q}\Delta \boldsymbol{K}_{\phi}^{T} \Delta \boldsymbol{K}_{r}^{T} \Delta \boldsymbol{K}_{v}^{T}) \in \mathcal{R}^{N_{\phi}\times N_{r}\times N_{v}}, \\
\boldsymbol{A}&=|\boldsymbol{A}|,  
\end{split}
\label{equ:12}
\end{equation}
where operator $|\boldsymbol{A}|$ calculates the absolute value of the complex numbers in $\boldsymbol{A}$. The $\text{argmax}$ in the algorithm is replaced by $\text{softmax}$ with reweighing coefficient $\sigma=1$ in Eq. (\ref{equ:12}). 
Denote $\boldsymbol{A}_{i} \in \mathcal{R}^{N_{\phi}}$, $\boldsymbol{A}_{j} \in \mathcal{R}^{N_{r}}$, and $\boldsymbol{A}_{k} \in \mathcal{R}^{N_{v}}$ as the 1D attention maps calculated from the 3D attention map by,
\begin{equation}
\begin{split}
\boldsymbol{A}_{i}&=\sum_{j=1}^{N_{r}}\sum_{k=1}^{N_{v}}[\boldsymbol{A}]_{i,j,k}, 
\boldsymbol{A}_{j}=\sum_{i=1}^{N_{\phi}}\sum_{k=1}^{N_{v}}[\boldsymbol{A}]_{i,j,k}, \\
\boldsymbol{A}_{k}&=\sum_{i=1}^{N_{\phi}}\sum_{j=1}^{N_{r}}[\boldsymbol{A}]_{i,j,k}.
\end{split}
\label{equ:13}
\end{equation}
Line 4 and line 5 in Algorithm \ref{alg:one} can be further modelled as multiplying 1D attention maps with values and adding the results to sets $\mathcal{T}$ and $\mathcal{M}$,
\begin{equation}
\begin{split}
\boldsymbol{F}_{attn}^{(1)}&=[\phi_{m}, r_{m}, v_{m}]=[\boldsymbol{A}_{i}\Delta \boldsymbol{V}_{\phi}^{(1)},  \boldsymbol{A}_{j}\Delta \boldsymbol{V}_{r}^{(1)}, \boldsymbol{A}_{k}\Delta \boldsymbol{V}_{v}^{(1)}], \\
\mathcal{T}_{k+1}&=\mathcal{T}_{k} \cup \boldsymbol{F}_{attn}^{(1)},    
\end{split}
\label{equ:14}
\end{equation}

\begin{equation}
\begin{split}
\boldsymbol{F}_{attn}^{(2)}&=\boldsymbol{\varphi}_{m}=\boldsymbol{a}(\phi_{i}) \otimes \boldsymbol{\rho}(r_{j}) \otimes \boldsymbol{\beta}(v_{k}) \\
&=\boldsymbol{A}_{i}\Delta \boldsymbol{V}_{\phi}^{(2)} \otimes \boldsymbol{A}_{j}\Delta \boldsymbol{V}_{r}^{(2)} \otimes \boldsymbol{A}_{k}\Delta \boldsymbol{V}_{v}^{(2)}, \\
\mathcal{M}_{k+1}&=\mathcal{M}_{k} \cup \boldsymbol{F}_{attn}^{(2)},      
\end{split}
\label{equ:15}
\end{equation}
where the definition of operator $\Delta$ is the same as Eq. (\ref{equ:9}). From Eq. (\ref{equ:12}) $\sim$ (\ref{equ:15}), line 3 $\sim$ line 5 in 3D-OMP algorithm consist of the attention layer in a transformer. 

To construct the FFN layer, we first define a matrix $\boldsymbol{M}_{m}=[\varphi_{1}, \varphi_{2}, \cdots, \varphi_{m}] \in \mathcal{C}^{KST_{s}\times m}$ from $\mathcal{M}_{m+1}$. We then rewrite the estimation problem in the line 6 of Algorithm \ref{alg:one} as 
\begin{equation}
\begin{split}
\hat{\boldsymbol{\gamma}}= \text{argmin}_{\boldsymbol{\gamma}} ||\tilde{\boldsymbol{Z}}-\boldsymbol{M}_{m}\boldsymbol{\gamma}|| 
\end{split}
\label{equ:16}
\end{equation}
The closed-form solution to Eq. (\ref{equ:16}) can be further given as, 
\begin{equation}
\begin{split}
\hat{\boldsymbol{\gamma}}= (\boldsymbol{M}_{m}^{H}\boldsymbol{M}_{m})^{-1}\boldsymbol{M}_{m}^{H}\tilde{\boldsymbol{Z}},
\end{split}
\label{equ:17}
\end{equation} 
With Eq. (\ref{equ:17}), the update of residual in the line 7 of Algorithm \ref{alg:one} is,
\begin{equation}
\begin{split}
\tilde{\boldsymbol{Z}}^{(m+1)}&= \tilde{\boldsymbol{Z}}-\boldsymbol{M}_{m}(\boldsymbol{M}_{m}^{H}\boldsymbol{M}_{m})^{-1}\boldsymbol{M}_{m}^{H}\tilde{\boldsymbol{m}}, \\
&=f(\boldsymbol{M}_{m}),
\end{split}
\label{equ:18}
\end{equation} 
where the tensor form of $\tilde{\boldsymbol{Z}}^{(m+1)}$ is the output of the $m$-th iteration. Obviously, $f$ consists of a series of linear transformations and non-linear activations based on $\boldsymbol{M}_{m}$. Therefore, we can safely write line 6 $\sim$ line 7 as an FFN layer in a Transformer, 
\begin{equation}
\boldsymbol{F}_{out}=\hat{\boldsymbol{Z}}^{(m+1)}=\text{FFN}(\mathcal{M}_{m+1}).
\label{equ:19}
\end{equation} 
We have now rewritten the 3D-OMP algorithm in the form of transformers. The details are summarized in Algorithm \ref{alg:two}. 

Algorithm \ref{alg:two} actually provides a new perspective of understanding transformers. From the algorithm, the attention layer finds the atom that maximally reduces the approximation error and the FFN layer subtracts the influence of the atom from the original signal for further processing. Algorithm \ref{alg:two} also provides new insights on transformer designs. For example, we could implement efficient 3D attention by designing three sets of keys and values, each set applying to one dimension of 3D input features.

Note that once a target's information is added into $\mathcal{T}$, it remains unchanged until the end. From the perspective of transformers, this property can be achieved by defining $\boldsymbol{Q}=\boldsymbol{K}=\boldsymbol{V}$ and setting the FFN as an identity mapping for $\mathcal{T}$. We omit this part in Algorithm \ref{alg:two} for simplicity.

\SetKwInput{KwInput}{Input}                
\SetKwInput{KwOutput}{Output}
\SetKwInput{KwInitial}{Initialization}
\SetKwComment{Comment}{/* }{ */}
\begin{algorithm}[t!]
\caption{Transformer Version of 3D-OMP for Multi-target Detection}\label{alg:two}

\KwInput{Observation $\boldsymbol{\hat{Z}}$ in (\ref{equ:6}), termination threshold $\delta$.}
\KwOutput{Set $\mathcal{T}=\{\phi_{m}, r_{m}, v_{m}\}_{m=1}^{M}$}
\KwInitial{$m=1$, $\boldsymbol{Q}^{(1)}=\boldsymbol{\hat{Z}}$, $\boldsymbol{K}_{\phi}=\boldsymbol{\Phi}_{\phi}^{H}$, $\boldsymbol{K}_{r}=\boldsymbol{\Phi}_{r}^{H}$,
$\boldsymbol{K}_{v}=\boldsymbol{\Phi}_{v}^{H}$,
$\boldsymbol{V}_{\phi}^{(1)}=\boldsymbol{G}_{\phi}$, 
$\boldsymbol{V}_{r}^{(1)}=\boldsymbol{G}_{r}$, 
$\boldsymbol{V}_{v}^{(1)}=\boldsymbol{G}_{v}$,
$\boldsymbol{V}_{\phi}^{(2)}=\boldsymbol{\Phi}_{\phi}$, 
$\boldsymbol{V}_{r}^{(2)}=\boldsymbol{\Phi}_{r}$, 
$\boldsymbol{V}_{v}^{(2)}=\boldsymbol{\Phi}_{v}$, 
}
$\boldsymbol{A}=\text{softmax}(\sigma \boldsymbol{Q}\Delta \boldsymbol{K}_{\phi}^{T} \Delta \boldsymbol{K}_{r}^{T} \Delta \boldsymbol{K}_{v}^{T}) \in \mathcal{R}^{N_{\phi}\times N_{r}\times N_{v}}$

\While{$\text{max} \,|\boldsymbol{A}|>\delta$}{
  \Comment*[h]{the attention-layer of the $m$-th iteration}
  
  Calculate the attention map 
  $\boldsymbol{A}=\text{softmax}(\sigma \boldsymbol{Q}^{(m)}\Delta \boldsymbol{K}_{\phi}^{T} \Delta \boldsymbol{K}_{r}^{T} \Delta \boldsymbol{K}_{v}^{T})$; 
  $\boldsymbol{A}=|\boldsymbol{A}|$

  Update outputs $\boldsymbol{F}_{attn}^{(1)}=[\phi_{m}, r_{m}, v_{m}]=[\boldsymbol{A}_{i}\Delta \boldsymbol{V}_{\phi}^{(1)},  \boldsymbol{A}_{j}\Delta \boldsymbol{V}_{r}^{(1)}, \boldsymbol{A}_{k}\Delta \boldsymbol{V}_{v}^{(1)}]$;
  $\boldsymbol{F}_{attn}^{(2)}=\boldsymbol{\varphi}_{m}=\boldsymbol{a}(\phi_{i}) \otimes \boldsymbol{\rho}(r_{j}) \otimes \boldsymbol{\beta}(v_{k})=\boldsymbol{A}_{i}\Delta \boldsymbol{V}_{\phi}^{(2)} \otimes \boldsymbol{A}_{j}\Delta \boldsymbol{V}_{r}^{(2)} \otimes \boldsymbol{A}_{k}\Delta \boldsymbol{V}_{v}^{(2)}$
  
  Update target set
  $\mathcal{T}_{m+1}=\mathcal{T}_{m} \cup \boldsymbol{F}_{attn}^{(1)}$

  Update atom set
  $\mathcal{M}_{m+1}=\mathcal{M}_{m} \cup \boldsymbol{F}_{attn}^{(2)}$

  \Comment*[h]{the FFN layer of the $m$-th iteration}

  $\boldsymbol{M}_{m}=[\varphi_{1}, \varphi_{2}, \cdots, \varphi_{m}] \in \mathcal{C}^{KST_{s}\times m}$

  $F_{out}= \tilde{\boldsymbol{Z}}-\boldsymbol{M}_{m}(\boldsymbol{M}_{m}^{H}\boldsymbol{M}_{m})^{-1}\boldsymbol{M}_{m}^{H}\tilde{\boldsymbol{Z}}=FFN(\boldsymbol{M}_{m})$
  
  \Comment*[h]{next iteration}
  
  $m=m+1$
  
  $\boldsymbol{Q}^{(m)}=\boldsymbol{F}_{out}$
  
}

\end{algorithm}

\section{3D-OMP-Transformer}
\begin{figure}[!ht]
\centering
\includegraphics[scale=0.75]{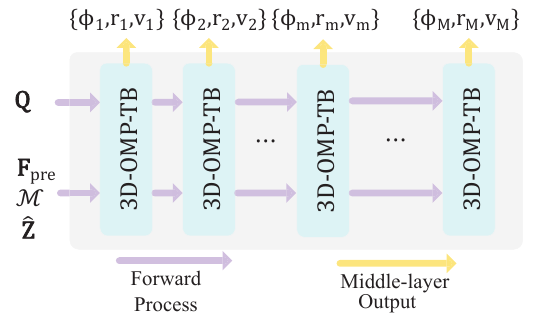}
\caption{The overall architecture of 3D-OMP-Transformer.}
\label{fig:3D-OMP-Transformer}
\end{figure}

\begin{figure*}[!ht]
\centering
\includegraphics[scale=0.65]{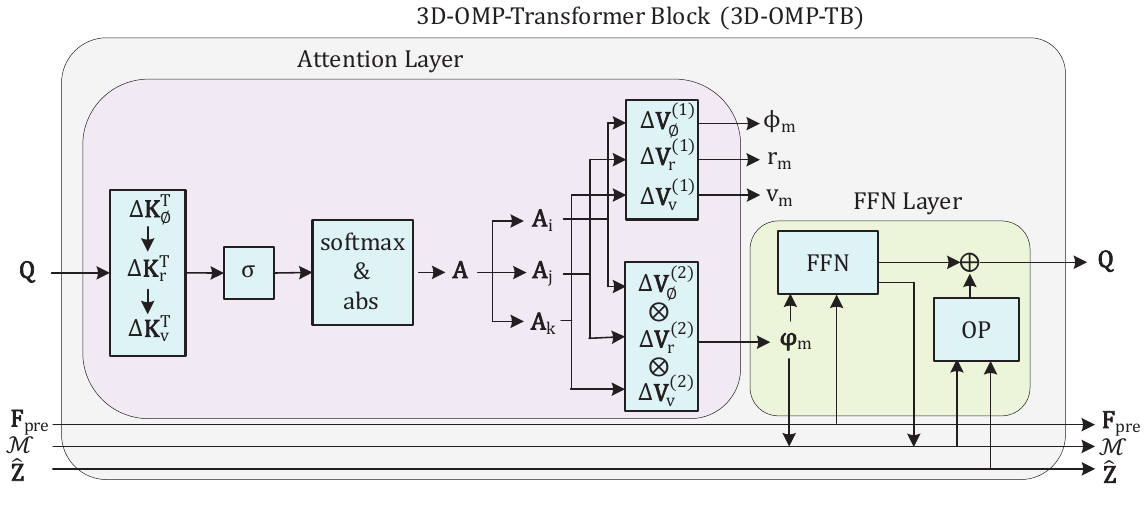}
\caption{The details of 3D-OMP Transformer block.}
\label{fig:3D-OMP-Transformer-block}
\end{figure*}
Although the operation of 3D-OMP algorithm is analogous to the transformers, 3D-OMP is designed with handcrafted parameters. Without learnable parameters, 3D-OMP algorithm lacks the ability to learn from data like other transformers. To address this issue and further improve the performance of 3D-OMP, we propose a 3D-OMP-Transformer, inspired by earlier success on deep unfolding networks \cite{gregor2010learning,sun2016deep} and differentiable OMP \cite{mateos2023model}. 

\subsection{Architecture}
The overall architecture of 3D-OMP-Transformer is shown in Fig. \ref{fig:3D-OMP-Transformer}. From the figure, the 3D-OMP-Transformer takes $\boldsymbol{\hat{Z}}$, $\boldsymbol{F}_{pre}=\boldsymbol{\hat{Z}}$, $\mathcal{M}=\emptyset$, and $\boldsymbol{Q}=\boldsymbol{\hat{Z}}$ as inputs and feeds the inputs into $M$ 3D-OMP-Transformer blocks (3D-OMP-TB). Each 3D-OMP-TB gives an estimation of one target. The estimated results are used to train the network in an end-to-end manner.

We now explain the detailed architecture of 3D-OMP-TB. In Fig. \ref{fig:3D-OMP-Transformer-block}, each 3D-OMP-TB is composed of one attention layer and one FFN layer. The attention layer takes $\boldsymbol{Q}$ as the input and calculates a 3D attention map $A$ by,
\begin{equation}
\boldsymbol{A}=\text{softmax}(\sigma \boldsymbol{Q}\Delta \boldsymbol{K}_{\phi}^{T} \Delta \boldsymbol{K}_{r}^{T} \Delta \boldsymbol{K}_{v}^{T}),
\boldsymbol{A}=|\boldsymbol{A}|,
\label{equ:20}
\end{equation} 
where $\boldsymbol{K}_{\phi}$, $\boldsymbol{K}_{r}$, $\boldsymbol{K}_{v}$, $\sigma$ are learnable parameters. We hope the network will learn to adjust the 3D-OMP algorithm rather than learning from scratch. To achieve this goal, we set,
\begin{equation}
\begin{split}
  \boldsymbol{K}_{\phi}&=\lambda_{ \boldsymbol{K}_{\phi}} \boldsymbol{\Phi}_{\phi}^{H}+\mu_{ \boldsymbol{K}_{\phi}} \boldsymbol{W}_{\boldsymbol{K}_{\phi}}, \\
  \boldsymbol{K}_{r}&=\lambda_{\boldsymbol{K}_{r}} \boldsymbol{\Phi}_{r}^{H}+\mu_{\boldsymbol{K}_{r}} \boldsymbol{W}_{\boldsymbol{K}_{r}}, \\
  \boldsymbol{K}_{v}&=\lambda_{\boldsymbol{K}_{v}} \boldsymbol{\Phi}_{v}^{H}+\mu_{\boldsymbol{K}_{v}} \boldsymbol{W}_{\boldsymbol{K}_{v}}, 
\end{split}
\label{equ:21}
\end{equation} 
where $\lambda_{ \boldsymbol{K}_{\phi}}$, $\mu_{ \boldsymbol{K}_{\phi}}$, $\lambda_{ \boldsymbol{K}_{r}}$, $\mu_{ \boldsymbol{K}_{r}}$, $\lambda_{ \boldsymbol{K}_{v}}$, $\mu_{ \boldsymbol{K}_{v}}$ are learnable scalars, and $\boldsymbol{W}_{\boldsymbol{K}_{\phi}}$, $\boldsymbol{W}_{\boldsymbol{K}_{r}}$, and $\boldsymbol{W}_{\boldsymbol{K}_{v}}$ are learnable matrices that have the same shapes as $\boldsymbol{\Phi}_{\phi}^{H}$, $\boldsymbol{\Phi}_{r}^{H}$, and $\boldsymbol{\Phi}_{v}^{H}$, respectively. During the parameter initialization process, we set $\lambda=1$, $\mu=0$, $\sigma=1$, so that Eq. (\ref{equ:20}) equals Eq. (\ref{equ:12}). In this way, the network will generate the same $\boldsymbol{A}$ as the 3D-OMP algorithm. But Eq. (\ref{equ:20}) will gradually improve the calculation of $\boldsymbol{A}$ based on the training. 

After obtaining $\boldsymbol{A}$, the attention layer will calculate the 1D attention maps $\boldsymbol{A}_{i}$, $\boldsymbol{A}_{j}$, and $\boldsymbol{A}_{k}$ by Eq. (\ref{equ:13}), and then use 1D attention maps to calculate the target's information $\{\phi_{m}, r_{m}, v_{m}\}$ and the atom of the same target, $\boldsymbol{\varphi}_{m}$, by,
\begin{equation}
[\phi_{m}, r_{m}, v_{m}]=[\boldsymbol{A}_{i}\Delta \boldsymbol{V}_{\phi}^{(1)},  \boldsymbol{A}_{j}\Delta \boldsymbol{V}_{r}^{(1)}, \boldsymbol{A}_{k}\Delta \boldsymbol{V}_{v}^{(1)}]
\label{equ:22}
\end{equation} 
and
\begin{equation}
\boldsymbol{\varphi}_{m}=\boldsymbol{A}_{i}\Delta \boldsymbol{V}_{\phi}^{(2)} \otimes \boldsymbol{A}_{j}\Delta \boldsymbol{V}_{r}^{(2)} \otimes \boldsymbol{A}_{k}\Delta \boldsymbol{V}_{v}^{(2)}
\label{equ:23}
\end{equation} 
Similarly, we set
\begin{equation}
\begin{split}
  \boldsymbol{V}_{\phi}^{(1)}&=\lambda_{ \boldsymbol{V}_{\phi}^{(1)}} \boldsymbol{G}_{\phi}+\mu_{ \boldsymbol{V}_{\phi}^{(1)}} \boldsymbol{W}_{\boldsymbol{V}_{\phi}^{(1)}}, \\
  \boldsymbol{V}_{r}^{(1)}&=\lambda_{ \boldsymbol{V}_{r}^{(1)}} \boldsymbol{G}_{r}+\mu_{ \boldsymbol{V}_{r}^{(1)}} \boldsymbol{W}_{\boldsymbol{V}_{r}^{(1)}}, \\
  \boldsymbol{V}_{v}^{(1)}&=\lambda_{ \boldsymbol{V}_{v}^{(1)}} \boldsymbol{G}_{v}+\mu_{ \boldsymbol{V}_{v}^{(1)}} \boldsymbol{W}_{\boldsymbol{V}_{v}^{(1)}}, \\
  \boldsymbol{V}_{\phi}^{(2)}&=\lambda_{ \boldsymbol{V}_{\phi}^{(2)}} \boldsymbol{\Phi}_{\phi}+\mu_{ \boldsymbol{V}_{\phi}^{(2)}} \boldsymbol{W}_{\boldsymbol{V}_{\phi}^{(2)}}, \\
  \boldsymbol{V}_{r}^{(2)}&=\lambda_{ \boldsymbol{V}_{r}^{(2)}} \boldsymbol{\Phi}_{r}+\mu_{ \boldsymbol{V}_{r}^{(2)}} \boldsymbol{W}_{\boldsymbol{V}_{r}^{(2)}}, \\
  \boldsymbol{V}_{v}^{(2)}&=\lambda_{ \boldsymbol{V}_{v}^{(2)}} \boldsymbol{\Phi}_{v}+\mu_{ \boldsymbol{V}_{v}^{(2)}} \boldsymbol{W}_{\boldsymbol{V}_{v}^{(2)}},
\end{split}
\label{equ:24}
\end{equation} 
where $\lambda_{\boldsymbol{V}_{\phi}^{(1)}}$, $\mu_{ \boldsymbol{V}_{\phi}^{(1)}}$, $\lambda_{ \boldsymbol{V}_{r}^{(1)}}$, $\mu_{\boldsymbol{V}_{r}^{(1)}}$, $\lambda_{\boldsymbol{V}_{v}^{(1)}}$, $\mu_{ \boldsymbol{V}_{v}^{(1)}}$, $\lambda_{\boldsymbol{V}_{\phi}^{(2)}}$, $\mu_{ \boldsymbol{V}_{\phi}^{(2)}}$, $\lambda_{ \boldsymbol{V}_{r}^{(2)}}$, $\mu_{\boldsymbol{V}_{r}^{(2)}}$, $\lambda_{\boldsymbol{V}_{v}^{(2)}}$, $\mu_{ \boldsymbol{V}_{v}^{(2)}}$ are learnable scalars, and $\boldsymbol{W}_{\boldsymbol{V}_{\phi}^{(1)}}$, $\boldsymbol{W}_{\boldsymbol{V}_{r}^{(1)}}$, $\boldsymbol{W}_{\boldsymbol{V}_{v}^{(1)}}$,
$\boldsymbol{W}_{\boldsymbol{V}_{\phi}^{(2)}}$, $\boldsymbol{W}_{\boldsymbol{V}_{r}^{(2)}}$, $\boldsymbol{W}_{\boldsymbol{V}_{v}^{(2)}}$ are learnable matrices. At the beginning, we set $\lambda=1$, $\mu=0$, $\sigma=1$, so that Eq. (\ref{equ:22}) equals Eq. (\ref{equ:14}), Eq. (\ref{equ:23}) equals Eq. (\ref{equ:15}); But Eq. (\ref{equ:22}) and Eq. (\ref{equ:23}) will gradually perform differently based on the training.  After the attention layer, $\{\phi_{m}, r_{m}, v_{m}\}$ will be used for calculating the trainng loss and $\boldsymbol{\varphi}_{m}$ will be added into $\mathcal{M}$ and used for the next 3D-OMP-TB. 

On the other hand, the FFN layer is composed of an FFN and an orthogonal projection (OP). The FFN consists of one 1x1 convolution layer (Conv), one 3x3 Conv, and one 1x1 Conv, with GELU activation layers in between. The FFN takes $\boldsymbol{\varphi}_{m}$ and $\boldsymbol{F}_{pre}$ as inputs and generates an output feature map $\boldsymbol{F}_{out}^{(1)}$ and an updated feature map $\boldsymbol{F}_{pre}$. The process can be modelled as,
\begin{equation}
\boldsymbol{F}_{out}^{(1)}, \boldsymbol{F}_{pre}= \text{FFN
}(\boldsymbol{\varphi}_{m}, \boldsymbol{F}_{pre})\label{equ:25}
\end{equation}
$\boldsymbol{F}_{pre}$ will be used for the FFN in the next 3D-OMP-TB. We also hope the FFN layer performs the same as the 3D-OMP algorithm at the beginning of training but will gradually learn from data. To achieve this goal, we add an OP inside the FFN layer. The OP utilizes $\mathcal{M}$ and $\boldsymbol{\hat{Z}}$ and calculates an output $\boldsymbol{F}_{out}^{(2)}$ based on Eq. (\ref{equ:18}). We define the output from the FFN layer as,
\begin{equation}
\boldsymbol{F}_{out}=\lambda_{\boldsymbol{F}_{out}} \boldsymbol{F}_{out}^{(2)}+\mu_{\boldsymbol{F}_{out}} \boldsymbol{F}_{out}^{(1)},
\label{equ:26}
\end{equation}
where $\lambda_{\boldsymbol{F}_{out}}$ and $\mu_{\boldsymbol{F}_{out}}$ are learnable scalars. The initialization values of $\lambda_{\boldsymbol{F}_{out}}$ and $\mu_{\boldsymbol{F}_{out}}$ are $1$ and $0$, respectively. Therefore, the network performs the same as 3D-OMP algorithm at the beginning. 

At last, $\boldsymbol{F}_{out}$ will be used as $\boldsymbol{Q}$ for the next 3D-OMP-TB. Updated $\mathcal{M}$, $\boldsymbol{F}_{pre}$ along with the original signal $\boldsymbol{\hat{Z}}$ are also sent to the next 3D-OMP-TB.
\subsection{Training Loss}
In this subsection, we will introduce the training loss for the proposed 3D-OMP-Transformer. Denote the true angle set, range set, and velocity set to be $\mathcal{A}=\{\phi_{m}\}_{m=1}^{M}$, $\mathcal{R}=\{r_{m}\}_{m=1}^{M}$, $\mathcal{V}=\{v_{m}\}_{m=1}^{M}$, respectively, and the estimated angle set, range set, and velocity set to be $\hat{\mathcal{A}}=\{\hat{\phi}_{m}\}_{m=1}^{M}$, $\hat{\mathcal{R}}=\{\hat{r}_{m}\}_{m=1}^{M}$, $\hat{\mathcal{V}}=\{\hat{v}_{m}\}_{m=1}^{M}$, respectively. We need to define one-to-one mappings between the elements in $\mathcal{A}$ and $\hat{\mathcal{A}}$, $\mathcal{R}$ and $\hat{\mathcal{R}}$, $\mathcal{V}$ and $\hat{\mathcal{V}}$ before calculating the training loss. To decide the mapping relations, we use the first-match-first-out criteria. Take $\mathcal{A}$ and $\hat{\mathcal{A}}$ as an example. We first find the closest elements in $\mathcal{A}$ and $\hat{\mathcal{A}}$ by,
\begin{equation}
(i,j)=\text{argmin}_{i,j}|[\mathcal{A}]_{i}-[\hat{\mathcal{A}}]_{j}|,
\label{equ:27}
\end{equation}
where $[\mathcal{A}]_{i}$ and $[\hat{\mathcal{A}}]_{j}$ are the $i$-th element and the $j$-th element in $\mathcal{A}$ and $\hat{\mathcal{A}}$, respectively. We then add $(i,j)$ into a set $\mathcal{G}_{\mathcal{A}}$ and delete $[\mathcal{A}]_{i}$ and $[\hat{\mathcal{A}}]_{j}$ from $\mathcal{A}$ and $\hat{\mathcal{A}}$, respectively. We repeat this process until $\mathcal{A}$ and $\hat{\mathcal{A}}$ are empty. Similarly, we obtain $\mathcal{G}_{\mathcal{R}}$, $\mathcal{G}_{\mathcal{V}}$ for range sets and velocity sets, respectively.

After deciding the mapping relations, we use mean absolute error (MAE) for training. The MAE for angles can be calculated as,
\begin{equation}
l_{\mathcal{A}}=\frac{1}{M}\sum_{i=1}^{M}\sum_{j=1}^{M} \delta_{(i,j)\in \mathcal{G}_{\mathcal{A}}} |[\mathcal{A}]_{i}-[\hat{\mathcal{A}}]_{j}|,
\label{equ:28}
\end{equation}
where $\delta_{(i,j)\in \mathcal{G}_{\mathcal{A}}}=1$ only if the relation $(i,j)$ is stored in $\mathcal{G}_{\mathcal{A}}$. We can calculate the MAE for ranges and velocities in a similar way,
\begin{equation}
\begin{split}
    l_{\mathcal{R}}&=\frac{1}{M}\sum_{i=1}^{M}\sum_{j=1}^{M} \delta_{(i,j)\in \mathcal{G}_{\mathcal{R}}} |[\mathcal{R}]_{i}-[\hat{\mathcal{R}}]_{j}|, \\
    l_{\mathcal{V}}&=\frac{1}{M}\sum_{i=1}^{M}\sum_{j=1}^{M} \delta_{(i,j)\in \mathcal{G}_{\mathcal{V}}} |[\mathcal{V}]_{i}-[\hat{\mathcal{V}}]_{j}|.
\end{split}
\label{equ:29}
\end{equation}
The final training loss is,
\begin{equation}
l=l_{\mathcal{A}}+l_{\mathcal{R}}+l_{\mathcal{V}}
\label{equ:30}
\end{equation}
\subsection{Dynamic Dictionary Range}
The region of targets may change. In this paper, we assume $\phi_{min}$ will change randomly, but $\phi_{max}-\phi_{min}$ keeps constant for simplicity. The 3D-OMP algorithm can deal with the change by setting the dictionaries and grids according to the region of targets with Eq. (\ref{equ:7}), but the 3D-OMP-Transformer with region-agnostic learnable parameters cannot. To address this issue, the learnable matrices used in the 3D-OMP-Transformer, i.e., $\boldsymbol{W}$ in Eq.(\ref{equ:21}) and Eq.(\ref{equ:24}), should also change dynamically. We will explain how to adjust the angle-related parameters, i.e., $\boldsymbol{W}_{\boldsymbol{K}_{\phi}}$, $\boldsymbol{W}_{\boldsymbol{V}_{\phi}^{(1)}}$, and $\boldsymbol{W}_{\boldsymbol{V}_{\phi}^{(2)}}$, according to the values of $\phi_{min}$.  

We first analyze the effect of angle changes on the angle dictionary in Eq. (\ref{equ:7}). From Eq. (\ref{equ:2}), if $\phi_{b}=\phi_{a}+ \phi$ and we assume $\sin{\phi_{a}}\simeq \frac{2}{\pi}\phi_{a}$, for $-\frac{\pi}{2}\leq \phi_{a} \leq \frac{\pi}{2}$, we have $\boldsymbol{a}(\phi_{b}) \simeq \boldsymbol{a}(\phi_{a})\odot\boldsymbol{a}(\phi)$. Therefore, if $\phi_{min}$ changes from $\phi_{a}$ to $\phi_{a}+ \phi$, the angle dictionary also changes from $\boldsymbol{\Phi}_{\phi}$ to $\boldsymbol{\Phi}_{\phi}\odot\boldsymbol{a}( \phi)$. This means we can adjust the dictionary by multiplying it with a complex vector to deal with the change of $\phi_{min}$. 

We then extend the same idea to learnable weights. To cope with the change of $\phi_{min}$, 3D-OMP-Transformer takes $\phi_{min}$ as inputs and generates $\boldsymbol{h}_{\boldsymbol{W}_{\boldsymbol{K}_{\phi}}}\in \mathcal{C}^{K\times 1}$, $h_{\boldsymbol{W}_{\boldsymbol{V}_{\phi}^{1}}}\in \mathcal{C}^{1}$, and  $\boldsymbol{h}_{\boldsymbol{W}_{\boldsymbol{V}_{\phi}^{2}}}\in \mathcal{C}^{K\times 1}$ by three sets of three-layer fully-connected networks (FCN). The 3D-OMP-Transformer then calculates the angle-related keys and values by,
\begin{equation}
\begin{split}
 \boldsymbol{K}_{\phi}&=\lambda_{\boldsymbol{K}_{\phi}} \boldsymbol{\Phi}_{\phi}^{H}+\mu_{\boldsymbol{K}_{\phi}}(\boldsymbol{h}_{\boldsymbol{W}_{\boldsymbol{K}_{\phi}}}\odot\boldsymbol{W}_{\boldsymbol{K}_{\phi}}), \\ \boldsymbol{V}_{\phi}^{(1)}&=\lambda_{\boldsymbol{V}_{\phi}^{(1)}} \boldsymbol{G}_{\phi}+\mu_{\boldsymbol{V}_{\phi}^{(1)}}(h_{\boldsymbol{W}_{\boldsymbol{V}_{\phi}^{(1)}}}+\boldsymbol{W}_{\boldsymbol{V}_{\phi}^{(1)}}), \\
\boldsymbol{V}_{\phi}^{(2)}&=\lambda_{\boldsymbol{V}_{\phi}^{(2)}} \boldsymbol{\Phi}_{\phi}+\mu_{\boldsymbol{V}_{\phi}^{(2)}}(\boldsymbol{h}_{\boldsymbol{W}_{\boldsymbol{V}_{\phi}^{(2)}}}\odot\boldsymbol{W}_{\boldsymbol{V}_{\phi}^{(2)}}).
\end{split}
\label{equ:31}
\end{equation}
The angle-related keys and values will be used in Eq. (\ref{equ:20}), Eq. (\ref{equ:22}), and Eq. (\ref{equ:23}). The other parts of the 3D-OMP-Transformer remain unchanged.

\section{Cascaded 3D-OMP-Transformer}
\begin{figure}[ht]
\centering
\includegraphics[scale=0.56]{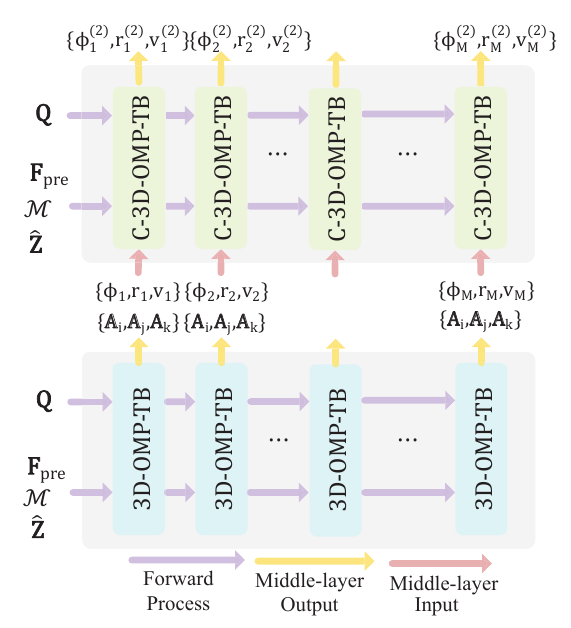}
\caption{The overall architecture of C-3D-OMP-Transformer.}
\label{fig:C-3D-OMP-Transformer}
\end{figure}

\begin{figure*}[ht]
\centering
\includegraphics[scale=0.65]{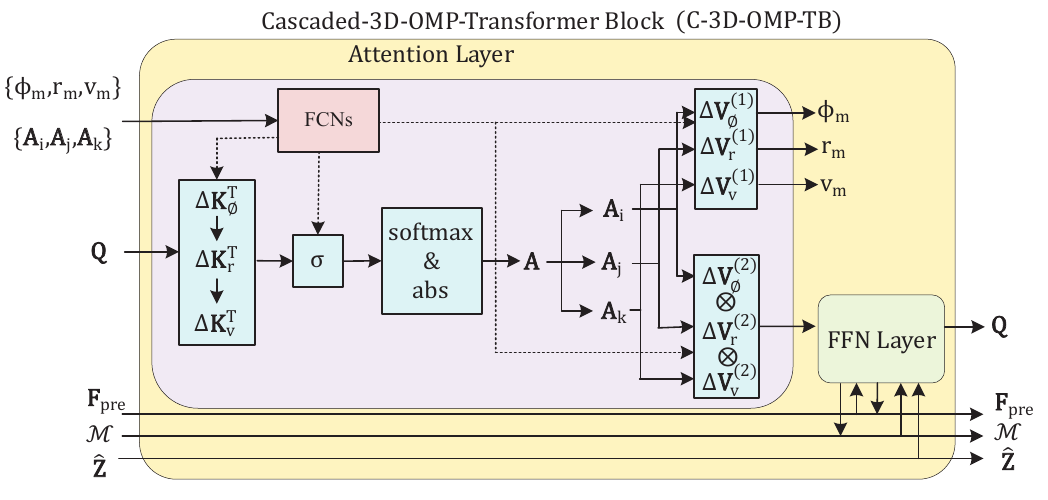}
\caption{The details of C-3D-OMP Transformer block.}
\label{fig:C-3D-OMP-Transformer-block}
\end{figure*}

The size of keys and values decides the resolution of dictionaries and grids, which will further determine the performance of the 3D-OMP-Transformer. However, the memory and computation costs in calculating 3D attention maps are increased with the size of keys. To address this issue, we further design a cascaded 3D-OMP transformer (C-3D-OMP-Transformer). The proposed C-3D-OMP-Transformer will use dynamic keys and values of small sizes to improve resolution. 

\subsection{Architecture}
The overall architecture of C-3D-OMP-Transformer is shown in Fig. \ref{fig:C-3D-OMP-Transformer}. From the figure, the C-3D-OMP-Transformer is built on top of the 3D-OMP-Transformer using the outputs of the latter and cascaded 3D-OMP-Transformer blocks (C-3D-OMP-TBs). The details of C-3D-OMP-TB are shown in Fig. \ref{fig:C-3D-OMP-Transformer-block}. Specifically, the C-3D-OMP-TB will feed the attention maps ($\boldsymbol{A}_{i}$, $\boldsymbol{A}_{j}$, $\boldsymbol{A}_{k}$) and the estimated results ($\Phi_{m}$, $r_{m}$, $v_{m}$) from 3D-OMP-TB into a set of FCNs and use the outputs of the FCNs to adjust the keys and values in the attention layer. The adjustment increases the resolution of grids and dictionaries with low costs. Hereafter, we will explain the adjustment process of $\boldsymbol{K}_{\phi}, \boldsymbol{V}_{\phi}^{(1)}, \boldsymbol{V}_{\phi}^{(2)}$. The other parameters, such as $\boldsymbol{K}_{r}, \boldsymbol{V}_{r}^{(1)}, \boldsymbol{V}_{r}^{(2)}$, $\boldsymbol{K}_{v}, \boldsymbol{V}_{v}^{(1)},\boldsymbol{V}_{v}^{(2)}$, are adjusted similarly.

We first determine the angle grids, $\boldsymbol{G}_{\phi}\in \mathcal{R}^{N_{\phi}^{'}\times 1}$, and angle dictionaries, $\boldsymbol{\Phi}_{\phi} \in \mathcal{C}^{K \times N_{\phi}^{'}}$, in the C-3D-OMP-TB, where $N_{\phi}^{'}<N_{\phi}$. Instead of using a fixed grid inside $[\phi_{min}, \phi_{max}]$, we let the network decide the optimal angle grid by considering $\boldsymbol{A}_{i}$ and $\phi_{m}$ and set the angle grid dynamically by,
\begin{equation}
\begin{split}
g_{min}, g_{max}&=\text{FCN}(\boldsymbol{A}_{i});  \\
\boldsymbol{G}_{\phi}&=(g_{max}-g_{min})*\boldsymbol{a}+g_{min}+\phi_{m},
\end{split}
\label{equ:32}
\end{equation}
where $g_{min}$, $g_{max}$ are two scalars generated from a three-layer FCN using $\boldsymbol{A}_{i}$, $\boldsymbol{a}\in \mathcal{R}^{N_{\phi}^{'}\times 1}$ is a fixed grid distributed uniformly from $-0.5$ to $0.5$. From Eq. (\ref{equ:32}), $g_{max}-g_{min}$ is a re-scaling weight of $\boldsymbol{a}$ and dynamically determines the angle resolution of $\boldsymbol{G}_{\phi}$. $g_{min}+\phi_{m}$ is an offset value and dynamically determines the central of $\boldsymbol{G}_{\phi}$. The network will learn to increase the angle resolution by adjusting $g_{max}-g_{min}$. $\boldsymbol{\Phi}_{\phi}$ can be further calculated from $\boldsymbol{G}_{\phi}$ with Eq. (\ref{equ:7})

Next, we make sure the learnable matrices ($\boldsymbol{W}_{\boldsymbol{K}_{\phi}}$, $\boldsymbol{W}_{\boldsymbol{V}_{\phi}^{(1)}}$, $\boldsymbol{W}_{\boldsymbol{V}_{\phi}^{(2)}}$) can change dynamically in a similar way as $\boldsymbol{G}_{\phi}$ and $\boldsymbol{\Phi}_{\phi}$ by considering $\boldsymbol{A}_{i}$ and $\phi_{m}$. As $\boldsymbol{W}_{\boldsymbol{V}_{\phi}^{(1)}}$ is used along with $\boldsymbol{G}_{\phi}$, we can impose the effect of re-scaling and offsetting on it by, 
\begin{equation}
\begin{split}
\tilde{g}_{min}, \tilde{g}_{max}&=\text{FCN}(\boldsymbol{A}_{i});  \\
\tilde{\boldsymbol{W}}_{\boldsymbol{V}_{\phi}^{(1)}} &=(\tilde{g}_{max}-\tilde{g}_{min})*\boldsymbol{W}_{\boldsymbol{V}_{\phi}^{(1)}}+\tilde{g}_{min}+\phi_{m},
\end{split}
\label{equ:33}
\end{equation}
where $\tilde{g}_{min}$, $\tilde{g}_{max}$ are two scalars generated from another three-layer FCN using $\boldsymbol{A}_{i}$. $\tilde{g}_{max}-\tilde{g}_{min}$ is a re-scaling weight and dynamically determines the angle range of $\tilde{\boldsymbol{W}}_{\boldsymbol{V}_{\phi}^{(1)}}$. $\tilde{g}_{min}+\phi_{m}$ is an offset value and dynamically determines the central of the $\tilde{\boldsymbol{W}}_{\boldsymbol{V}_{\phi}^{(1)}}$.

As $\boldsymbol{W}_{\boldsymbol{K}_{\phi}}$ and $\boldsymbol{W}_{\boldsymbol{V}_{\phi}^{(2)}}$ are used together with $\boldsymbol{\Phi}_{\phi}$, we first analyze the effect of re-scaling and offset on $\boldsymbol{\Phi}_{\phi}$. From Eq. (\ref{equ:2}), if $\phi_{b}=p\phi_{a}+ \phi$ and we assume $\sin{\phi_{a}}\simeq \frac{2}{\pi}\phi_{a}$, for $-\frac{\pi}{2}\leq \phi_{a} \leq \frac{\pi}{2}$, we have $\boldsymbol{a}(\phi_{b}) \simeq \boldsymbol{a}^{p}(\phi_{a}) \odot\boldsymbol{a}(\phi)$. Therefore, if a re-scaling parameter, $p$, and an offset, $\phi$, are applied to the angle grid, the angle dictionary also changes from $\boldsymbol{\Phi}_{\phi}$ to $\boldsymbol{\Phi}_{\phi}^{p}\odot\boldsymbol{a}( \phi)$. This means we can adjust $\boldsymbol{W}_{\boldsymbol{K}_{\phi}}$ and $\boldsymbol{W}_{\boldsymbol{V}_{\phi}^{(2)}}$ by first performing exponentiation to them and then multiplying them  with a complex vector. The process can be modelled as,
\begin{equation}
\begin{split}
g_{\boldsymbol{W}_{\boldsymbol{V}_{\phi}^{(2)}}}, \boldsymbol{h}_{\boldsymbol{W}_{\boldsymbol{V}_{\phi}^{(2)}}}&=\text{FCN}(\boldsymbol{A}_{i}),  \\
\Tilde{\boldsymbol{W}}_{\boldsymbol{V}_{\phi}^{(2)}}&=\boldsymbol{h}_{\boldsymbol{W}_{\boldsymbol{V}_{\phi}^{(2)}}}\odot(\boldsymbol{W}_{\boldsymbol{V}_{\phi}^{2}})^{g_{\boldsymbol{W}_{\boldsymbol{V}_{\phi}^{(2)}}}}, \\
g_{\boldsymbol{W}_{\boldsymbol{K}_{\phi}}}, \boldsymbol{h}_{\boldsymbol{W}_{\boldsymbol{K}_{\phi}}}&=\text{FCN}(\boldsymbol{A}_{i}),  \\
\tilde{\boldsymbol{W}}_{\boldsymbol{K}_{\phi}}&=\boldsymbol{h}_{\boldsymbol{W}_{\boldsymbol{K}_{\phi}}}\odot(\boldsymbol{W}_{\boldsymbol{K}_{\phi}})^{g_{\boldsymbol{W}_{\boldsymbol{K}_{\phi}}}}, 
\end{split}
\label{equ:34}
\end{equation}
where $g_{\boldsymbol{W}_{\boldsymbol{V}_{\phi}^{(2)}}}\in \mathcal{R}$, $g_{\boldsymbol{W}_{\boldsymbol{K}_{\phi}}}\in \mathcal{R}$ represent the influence of re-scaling and are estimated from $\boldsymbol{A}_{i}$ by a FCN. $\boldsymbol{h}_{\boldsymbol{W}_{\boldsymbol{V}_{\phi}^{(2)}}}\in \mathcal{C}^{K\times 1}$, $\boldsymbol{h}_{\boldsymbol{W}_{\boldsymbol{K}_{\phi}}} \in \mathcal{C}^{K\times 1}$ represent the influence of offset and are estimated from $\boldsymbol{A}_{i}$ by another FCN.

Finally, the angle-related keys and values are,
\begin{equation}
\begin{split}
\boldsymbol{K}_{\phi}&=\lambda_{\boldsymbol{K}_{\phi}} \boldsymbol{\Phi}_{\phi}^{H}+\mu_{\boldsymbol{K}_{\phi}}\tilde{\boldsymbol{W}}_{\boldsymbol{K}_{\phi}},  \\
\boldsymbol{V}_{\phi}^{(1)}&=\lambda_{\boldsymbol{V}_{\phi}^{(1)}} \boldsymbol{G}_{\phi}+\mu_{\boldsymbol{V}_{\phi}^{(1)}}\tilde{\boldsymbol{W}}_{\boldsymbol{V}_{\phi}^{(1)}}, \\
\boldsymbol{V}_{\phi}^{(2)}&=\lambda_{\boldsymbol{V}_{\phi}^{(2)}} \boldsymbol{\Phi}_{\phi}+\mu_{\boldsymbol{V}_{\phi}^{(2)}}
\tilde{\boldsymbol{W}}_{\boldsymbol{V}_{\phi}^{(2)}}.
\end{split}
\label{equ:35}
\end{equation}

Besides keys and values, the re-scaling parameter $\sigma$ in the softmax operation is also determined by a FCN.

\subsection{Training Loss}
In this subsection, we will describe the training loss for the C-3D-OMP-Transformer. Denote the estimated angle set, range set, and velocity set from a C-3D-OMP-Transformer are $\hat{\mathcal{A}}^{(2)}=\{\hat{\phi}_{m}^{(2)}\}_{m=1}^{M}$, $\hat{\mathcal{R}}^{(2)}=\{\hat{r}_{m}^{(2)}\}_{m=1}^{M}$, and $\hat{\mathcal{V}}^{(2)}=\{\hat{v}_{m}^{(2)}\}_{m=1}^{M}$, respectively. As the $m$-th C-3D-OMP-TB rely on the outputs of the $m$-th 3D-OMP-TB, we assume $\hat{\phi}_{m}$, $\hat{r}_{m}$, $\hat{v}_{m}$ and $\hat{\phi}_{m}^{(2)}$, $\hat{r}_{m}^{(2)}$, $\hat{v}_{m}^{(2)}$ are the estimation of the same target. Therefore, the C-3D-OMP-Transformer will reuse the mapping relation sets from 3D-OMP-Transformer, i.e., $\mathcal{G}_{A}$, $\mathcal{G}_{V}$, and $\mathcal{G}_{V}$. The MAE for angles, ranges, and velocities are,  
\begin{equation}
\begin{split}
    l_{\mathcal{A}^{(2)}}&=\frac{1}{M}\sum_{i=1}^{M}\sum_{j=1}^{M} \delta_{(i,j)\in \mathcal{G}_{\mathcal{A}}} |[\mathcal{A}]_{i}-[\hat{\mathcal{A}}^{(2)}]_{j}|, \\
    l_{\mathcal{R}^{(2)}}&=\frac{1}{M}\sum_{i=1}^{M}\sum_{j=1}^{M} \delta_{(i,j)\in \mathcal{G}_{\mathcal{R}}} |[\mathcal{R}]_{i}-[\hat{\mathcal{R}}^{(2)}]_{j}|, \\
    l_{\mathcal{V}^{(2)}}&=\frac{1}{M}\sum_{i=1}^{M}\sum_{j=1}^{M} \delta_{(i,j)\in \mathcal{G}_{\mathcal{V}}} |[\mathcal{V}]_{i}-[\hat{\mathcal{V}}^{(2)}]_{j}|,
\end{split}
\label{equ:36}
\end{equation}
respectively. The C-3D-OMP-Transformer is jointly trained with the 3D-OMP-Transformer; therefore, the total training loss is,
\begin{equation}
l=\frac{1}{2}(l_{\mathcal{A}}+l_{\mathcal{R}}+l_{\mathcal{V}}+l_{\mathcal{A}^{(2)}}+l_{\mathcal{R}^{(2)}}+l_{\mathcal{V}^{(2)}})
\label{equ:37}
\end{equation}

\section{Experiments}
In this section, we will first describe the default experimental settings and then provide the implementation details of the designed algorithms. Next, we will introduce the implementation details of the considered baselines, i.e., 1D-MUSIC+MF \cite{liu2020joint}, 2D-MUSIC+MF \cite{zheng2017super}, and 3D-OMP. Finally, we will compare the proposed algorithms with the baselines in the default settings, followed by some additional experiments with changed settings.


\subsection{Experimental Settings of the ISAC Scenario}
The parameters of the ISAC transceiver are, $K=16$, $S=128$, $T=10$. $f_{c}=60\,GHz$, and $\Delta_{f}=120 \,kHz$. $\Delta_{T}$ is set as $1/\Delta_{f}+1.5\,\mu s$, where $1.5\,\mu s$ is the time duration of the cyclic prefix. The number of targets, $M$, changes from $1\sim 5$. We set the parameters of targets as $r_{min}=0\,m$, $r_{max}=200\,m$, $v_{min}=0\,m/sec$, $v_{max}=42\,m/sec$, $\phi_{max}-\phi_{min}=40^{\circ}$, and $\phi_{min}$ uniformly distributed in $(-90^{\circ},50^{\circ})$. $\text{SNR}_{s}$ is set as $0\,dB$ or $10\,dB$.

\subsection{Implement Details of Proposed Methods} The parameters of the 3D-OMP-Transformer are, $N_{\phi}=360$, $N_{r}=300$, $N_{v}=60$. The parameters of the C-3D-OMP-Transformer are, $N_{\phi}^{'}=100$, $N_{r}^{'}=100$, $N_{v}^{'}=40$, where $N_{r}^{'}$ and $N_{v}^{'}$ are the sizes of range grids and velocity grids in the C-3D-OMP-Transformer. 

We train different transformers for different $M$. The 3D-OMP-Transformer is trained with a batch size of $8$ for $100,000$ rounds. The C-3D-OMP-Transformer is trained based on the pre-trained 3D-OMP-Transformer for another $100,000$ rounds. Each training sample is generated from Eq. (\ref{equ:6}) by randomly sampling targets' locations and channel noises. Both networks are trained by Adam optimizer with a $10^{-4}$ learning rate. 

We also generate testing datasets of size $10,000$ under each experimental setting for performance evaluation. We use the MAE of angles, ranges, and velocities in Eq. (\ref{equ:28}) and Eq. (\ref{equ:29}) as the performance metrics for the 3D-OMP-Transformer. We use the MAE of angles, ranges, and velocities in Eq. (\ref{equ:36}) to evaluate the performance of the C-3D-OMP-Transformer.

\subsection{Benchmarks}
Here are the multi-target detection baselines and their implementation details. All the methods use the same testing datasets and performance metrics as the 3D-OMP-Transformer. 
\begin{itemize}
    \item \textbf{1D-MUSIC+MF}: We considered the two-stage estimation method in \cite{liu2020joint}, where the angle information is first estimated by the 1D multiple signal classification (MUSIC) algorithm and the range/velocity information is then estimated by matching filter (MF) algorithm. The steering vector for the 1D MUSIC is $\boldsymbol{a_{rx}(\phi)}$ and the measurements in the frequency-time domain are snapshots. Different from \cite{liu2020joint}, we directly calculate the MUSIC spectrum based on the modulation symbols, as in \cite{sit2012direction}. To be specific, we first reshape $\boldsymbol{\hat{Z}}$ into $\boldsymbol{A}$ with the shape of $(K, ST)$ and then calculate the singular value decomposition (SVD) of $\boldsymbol{R}=\boldsymbol{AA^{H}}$. Denote the SVD result to be $\boldsymbol{R}=\boldsymbol{FDH}$, where $\boldsymbol{F}=[\boldsymbol{F^{s}}, \boldsymbol{F^{n}}]\in \mathcal{R}^{K \times K}$, $\boldsymbol{F^{s}}\in \mathcal{R}^{K \times M}$, and $\boldsymbol{F^{n}}\in \mathcal{R}^{K \times K-M}$. $\boldsymbol{F^{n}}$ is then used to calculate the MUSIC spectrum $f_{\text{MUSIC}}(\phi)=1/||(\boldsymbol{F^{n}})^{H}\boldsymbol{a}_{rx}(\phi)||_{2}$. Next, we try to find $M$ peaks from $f_{\text{MUSIC}}$. If less than $M$ peaks are detected, we reuse the detected peaks. After estimating the targets' AoAs, MF in \cite{liu2020joint} is implemented to estimate ranges and velocities. 
    \item \textbf{2D-MUSIC+MF}: As $K$ is not much greater than $M$ and $\phi_{max}-\phi_{min}$ is restricted to a small range, some targets are indistinguishable in the angle domain, restricting the performance of 1D-MUSIC. To address this, we implement the 2D-MUSIC algorithm specified in \cite{zheng2017super} for the joint estimation of angles and ranges. Joint estimation helps detect targets with similar angles but different ranges. In this process, smoothness along the angle and range dimensions is applied first to solve the problem of insufficient snapshots in the time domain. Specifically, $\boldsymbol{\text{Vec}}(\boldsymbol{a}_{rx}(\phi)_{:\hat{K}} \otimes \boldsymbol{\rho}(r)_{:\hat{S}}) \in \mathcal{R}^{\hat{K}\hat{S}}$ is treated as the steering vector of 2D-MUSIC, where $\boldsymbol{a}_{rx}(\phi)_{:\hat{K}}$ and $\boldsymbol{\rho}(r)_{:\hat{S}}$ denote the first $\hat{K}$ and $\hat{S}$ elements of $\boldsymbol{a}_{rx}(\phi)$ and $\boldsymbol{\rho}(r)$, respectively; $\boldsymbol{\text{Vec}(\cdot)}$ denotes reshaping a matrix into a vector. We notice that $\frac{\hat{K}}{K}=1$ and $\frac{\hat{S}}{S}=0.6$ work best. After smoothing, the observation matrix $\boldsymbol{\hat{Z}}$ will be of shape $(\hat{K}\hat{S}, \lfloor(K/\hat{K})\rfloor \lfloor(S/\hat{S})\rfloor T)$, which will be used to construct the 2D-MUSIC spectrum. After determining $M$ targets' angles and ranges, MF is applied for velocity estimation. 
    \item \textbf{3D-OMP}: The original 3D-OMP in Algorithm \ref{alg:one} is also considered. As we assume the number of targets is pre-known, we fix the round of iteration of the 3D-OMP algorithm as $M$. The size of the grids is set the same as the 3D-OMP-Transformer.
    
\end{itemize}

Note that we will not use optimization-based methods or deep unfolding networks to solve Eq. (\ref{equ:8}) as the size of $\boldsymbol{D}$, $(16\times 128 \times 10) \times (360\times 300 \times 60)$, is extremely large. We will not compare with the network-based solutions as existing networks perform no better than 3D-OMP when hardware imperfection is not considered \cite{mateos2023model}.

\subsection{Experiment Results}
\begin{table*}[!ht]
\centering
\caption{The performance comparison of different methods when $SNR_{s}=10\,dB$. }
\begin{tabular}{|c|c|c|c|c|c|c|}
\hline
M                  & MAE    & 1D-MUSIC+MF & 2D-MUSIC+MF & $\quad\,$3D-OMP$\,\quad$ & 3D-OMP-Transformer & C-3D-OMP-Transformer \\ \hline
\multirow{3}{*}{1} & $\phi$ &  $0.00193^{\circ}$                                                       &   $0.00118^{\circ}$                                                      &  $0.00106^{\circ}$      &  $0.00078^{\circ}$                           &   $0.00079^{\circ}$                                                                \\ \cline{2-7} 
 & $r$    &  $0.312\, m$                      & $0.241\, m$                                                       &$0.201\, m$        & $0.114\, m$                                                               &    $0.0767\, m$                                                              \\ \cline{2-7} 
& $v$    &  $0.234\, m/s$                                                        &   $0.205\, m/s$                                                     &  $0.205\, m/s$      &  $0.074\, m/s$                                                              &  $0.077\, m/s$                                                               \\ \hline
\multirow{3}{*}{2} & $\phi$ &  $0.0077^{\circ}$                                                      &   $0.00257^{\circ}$                                                     &   $0.00196^{\circ}$      &   $0.00171^{\circ}$                                                            &   $0.00169^{\circ}$                                                               \\ \cline{2-7} 
                   & $r$    &   $10.308\, m$                                                      &           $0.383\, m$                                             &   $0.258\, m$     &    $0.192\, m$                                                           &    $0.129\, m$                                                             \\ \cline{2-7} 
                   & $v$    &    $2.345\, m/s$                                                     &      $0.429\, m/s$                                                   & $0.286\, m/s$        &     $0.177\, m/s$                                                          &    $0.175\, m/s$                                                              \\ \hline
\multirow{3}{*}{3} & $\phi$ &    $0.0383^{\circ}$                                                      &   $0.00506^{\circ}$                                                     &  $0.00326^{\circ}$      &   $0.00309^{\circ}$                                                            &   $0.00297^{\circ}$                                                               \\ \cline{2-7} 
                   & $r$    &    $18.504\, m$                                                    &         $0.855\, m$                                                &  $0.450\, m$      &    $0.395\, m$                                                           &     $0.325\, m$                                                             \\ \cline{2-7} 
                   & $v$    &    $4.109\, m/s$                                                    &      $0.698\, m/s$                                                   &  $0.393\, m/s$        &  $0.295\, m/s$                                                              &   $0.266\, m/s$                                                                \\ \hline
\multirow{3}{*}{4} & $\phi$ &  $0.0688^{\circ}$                                                      &   $0.00803^{\circ}$                                                      &  $0.00473^{\circ}$      &  $0.0044^{\circ}$                                                             &  $0.0043^{\circ}$                                                                \\ \cline{2-7} 
                   & $r$    &     $24.554\, m$                                                    &  $1.487\, m$                                                      & $0.646\, m$        & $0.605\, m$                                                               & $0.509\, m$                                                                 \\ \cline{2-7} 
                   & $v$    &   $5.361\, m/s$                                                      &        $0.974\, m/s$                                                &  $0.484\, m/s$      &    $0.394\, m/s$                                                           &  $0.353\, m/s$                                                               \\ \hline
\multirow{3}{*}{5} & $\phi$ &   $0.0908^{\circ}$                                                     &   $0.0111^{\circ}$                                                      &  $0.0061^{\circ}$       &   $0.0057^{\circ}$                                                            &    $0.0055^{\circ}$                                                              \\ \cline{2-7} 
                   & $r$    &   $28.84\, m$                                &  $2.190\, m$                                                      &  $0.955\, m$      &  $0.897\, m$                                                               &  $0.803\, m$                                                               \\ \cline{2-7} 
                   & $v$    &   $6.34\, m/s$                          &  $1.226\, m/s$                                                       &  $0.601\, m/s$      &    $0.511\, m/s$                                                           &   $0.458\, m/s$                                                                \\ \hline
\end{tabular}
\label{tab:1}
\end{table*}
We show the performance comparison of different methods when $SNR_{s}=10\, dB$ in Table. \ref{tab:1}. From the table, 1D-MUSIC+MF does not perform well in multi-target situations. As $M$ increases, some targets are too close in the angle domain to be detected using the 1D-MUSIC algorithm. By jointly considering angle and range information, undetectable targets become distinguishable in the 2D angle-range domain; therefore, 2D-MUSIC+MF performs better than 1D-MUSIC+MF. However, 2D-MUSIC+MF still performs worse than 3D-OMP, showing the necessity of joint estimation in the 3D domain. 

On the other hand, the proposed 3D-OMP-Transformer outperforms the original 3D-OMP-Transformer. By learning to refine the 3D-OMP algorithm, 3D-OMP-Transformer reduces the angle error by about $7\%\sim25\%$, the range error by about $7\%\sim40\%$, and the velocity error by about $15\%\sim60\%$. The proposed C-3D-OMP-Transformer further reduces the localization error of the 3D-OMP-Transformer by learning to design the keys/values dynamically. To be exact, C-3D-OMP-Transformer can reduce the error of 3D-OMP-Transformer by up to $6\%$ in the angle estimation, $33\%$ in the range estimation, and $11\%$ in the velocity estimation. Overall, C-3D-OMP-Transformer achieves the best performance in multi-target detection.

\begin{table*}[!ht]
\centering
\caption{The performance comparison of different methods when $SNR_{s}=0\,dB$. }
\begin{tabular}{|c|c|c|c|c|c|c|}
\hline
M                  & MAE    & 1D-MUSIC+MF & 2D-MUSIC+MF & $\quad\,$3D-OMP$\,\quad$ & 3D-OMP-Transformer & C-3D-OMP-Transformer \\ \hline
\multirow{3}{*}{1} & $\phi$ &  $0.00937^{\circ}$                                                       &   $0.00469^{\circ}$                                                      &  $0.00234^{\circ}$      &  $0.00211^{\circ}$                           &   $0.00212^{\circ}$                                                                \\ \cline{2-7} 
 & $r$    &  $1.233\, m$                      & $0.874\, m$                                                       &$0.308\, m$        & $0.211\, m$                                                               &    $0.198\, m$                                                              \\ \cline{2-7} 
& $v$    &  $0.562\, m/s$                                                        &   $0.438\, m/s$                                                     &  $0.325\, m/s$      &  $0.238\, m/s$                                                              &  $0.238\, m/s$                                                               \\ \hline
\multirow{3}{*}{2} & $\phi$ &  $0.0325^{\circ}$                                                      &   $0.0101^{\circ}$                                                     &   $0.00403^{\circ}$      &   $0.00365^{\circ}$                                                            &   $0.00361^{\circ}$                                                               \\ \cline{2-7} 
                   & $r$    &   $12.370\, m$                                                      &           $1.728\, m$                                             &   $0.494\, m$     &    $0.391\, m$                                                           &    $0.347\, m$                                                             \\ \cline{2-7} 
                   & $v$    &    $2.936\, m/s$                                                     &      $0.827\, m/s$                                                   & $0.460\, m/s$        &     $0.387\, m/s$                                                          &    $0.382\, m/s$                                                              \\ \hline
\multirow{3}{*}{3} & $\phi$ &    $0.0712^{\circ}$                                                      &   $0.0165^{\circ}$                                                     &  $0.00647^{\circ}$      &   $0.00598^{\circ}$                                                            &   $0.00594^{\circ}$                                                               \\ \cline{2-7} 
                   & $r$    &    $21.966\, m$                                                    &         $3.026\, m$                                                &  $0.826\, m$      &    $0.782\, m$                                                           &     $0.728\, m$                                                             \\ \cline{2-7} 
                   & $v$    &    $4.950\, m/s$                                                    &      $1.246\, m/s$                                                   &  $0.652\, m/s$        &  $0.562\, m/s$                                                              &   $0.546\, m/s$                                                                \\ \hline
\multirow{3}{*}{4} & $\phi$ &  $0.1022^{\circ}$                                                      &   $0.0235^{\circ}$                                                      &  $0.00817^{\circ}$      &  $0.00748^{\circ}$                                                             & $0.00737^{\circ}$                                                                \\ \cline{2-7} 
                   & $r$    &     $28.241\, m$                                                    &  $4.714\, m$                                                      & $1.143\, m$        & $1.050\, m$                                                               & $0.992\, m$                                                                \\ \cline{2-7} 
                   & $v$    &   $6.249\, m/s$                                                      &        $1.698\, m/s$                                                &  $0.790\, m/s$      &    $0.677\, m/s$                                                           &   $0.655\, m/s$                                                              \\ \hline
\multirow{3}{*}{5} & $\phi$ &   $0.123^{\circ}$                                                     &   $0.0286^{\circ}$                                                      &  $0.0099^{\circ}$       &   $0.0092^{\circ}$                                                            &  $0.0091^{\circ}$                                                               \\ \cline{2-7} 
                   & $r$    &   $33.487\, m$                                &  $5.907\, m$                                                      &  $1.481\, m$      &  $1.375\, m$                                                               &  $1.313\, m$                                                               \\ \cline{2-7} 
                   & $v$    &   $7.260\, m/s$                          &  $2.026\, m/s$                                                       &  $0.951\, m/s$      &    $0.824\, m/s$                                                           &  $0.790\, m/s$                                                               \\ \hline
\end{tabular}
\label{tab:2}
\end{table*}

Table. \ref{tab:2} compares different methods when $SNR_{s}=0\, dB$. Similarly, 3D-OMP achieves better performance than 1D-MUSIC+MF and 2D-MUSIC+MF. 3D-OMP-Transformer reduces the error of 3D-OMP by about $4\%\sim16\%$ in angle estimation, by about $5\%\sim32\%$ in range estimation, and by about $14\%\sim27\%$ in velocity estimation. C-3D-OMP-Transformer can reduce the error of 3D-OMP-Transformer by up to $1.4\%$, $36\%$, $4\%$ in angle estimation, range estimation, and velocity estimation, respectively.

\begin{figure}[!ht]
\centering
\includegraphics[scale=0.7]{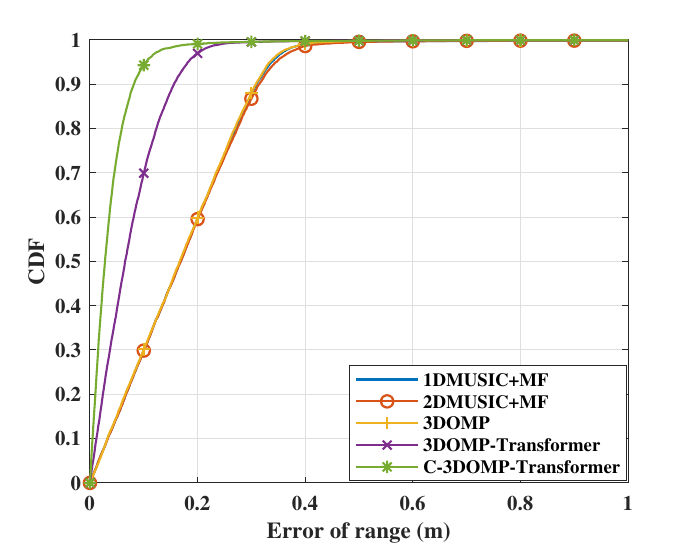}
\caption{The CDF of range error for different methods when $M=1$ and $SNR_{s}=10\, dB$.}
\label{fig:CDF-M=1-SNR10}
\end{figure}

\begin{figure}[!ht]
\centering
\includegraphics[scale=0.7]{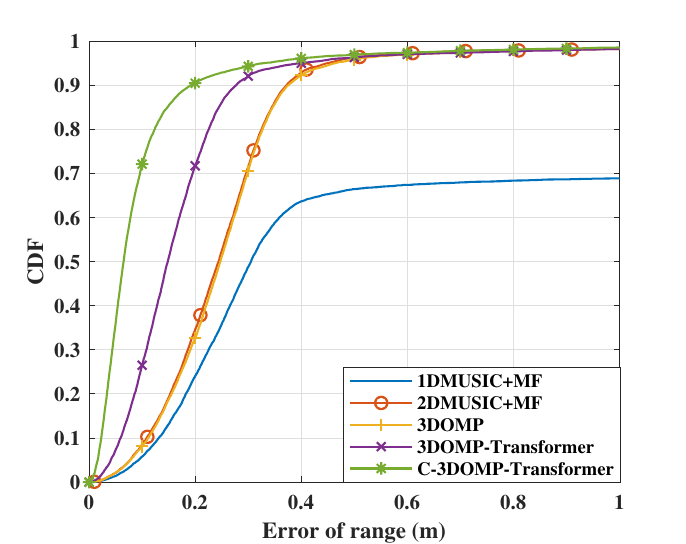}
\caption{The CDF of range error for the least match targets in different methods when $M=2$ and $SNR_{s}=10\, dB$.}
\label{fig:CDF-M=2-SNR10-leastmatch}
\end{figure}

Fig. \ref{fig:CDF-M=1-SNR10} shows the cumulative distribution function (CDF) of range errors for different methods when $M=1$ and $SNR_{s}=1\,dB$. The CDF calculates the probability of the range error being below a specific value in the $x$-axis. From the figure, 1D-MUSIC+MF, 2D-MUSIC+MF, and 3D-OMP have similar CDFs. The proposed 3D-OMP-Transformer and C-3D-OMP-Transformer have much better CDFs. 

Fig. \ref{fig:CDF-M=2-SNR10-leastmatch} shows the CDF of range error for the least match targets among $M$ targets for different methods when $M=2$ and $SNR_{s}=1\,dB$. The match relation is calculated from Eq. (\ref{equ:27}). From the figure, the proposed 3D-OMP-Transformer and C-3D-OMP-Transformer perform better than 1D-MUSIC+MF, 2D-MUSIC+MF, and 3D-OMP. 

\begin{figure*}[!ht]
\centering
\subfloat[CDF of the most match]{\includegraphics[width = 3in]{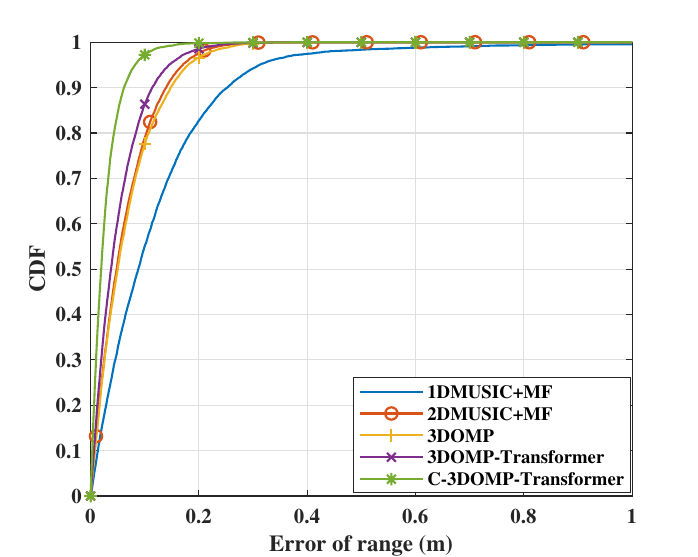}} 
\subfloat[CDF of the 2nd match]{\includegraphics[width = 3in]{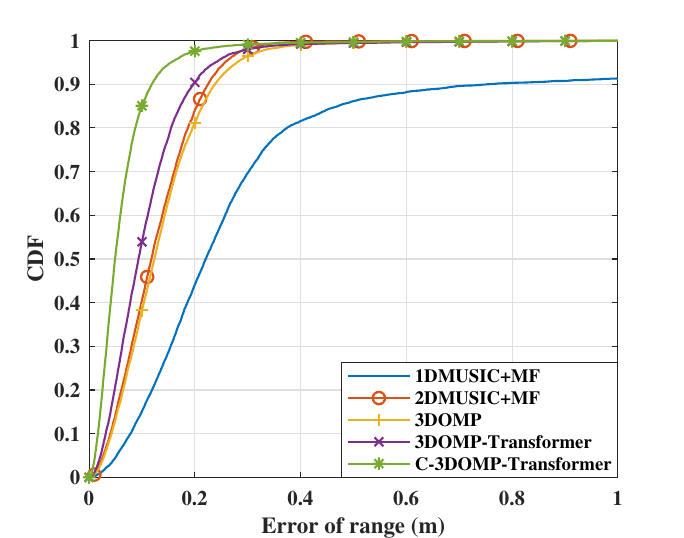}}\\
\subfloat[CDF of the 3rd match]{\includegraphics[width = 3in]{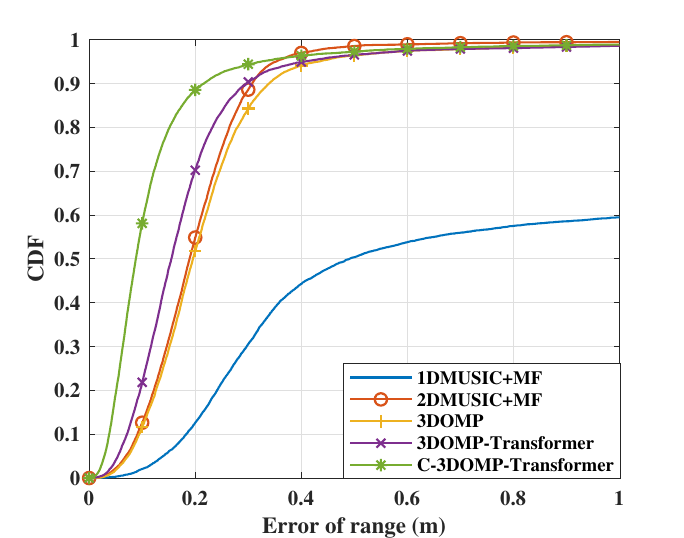}}
\subfloat[CDF of the 4th match]{\includegraphics[width = 3in]{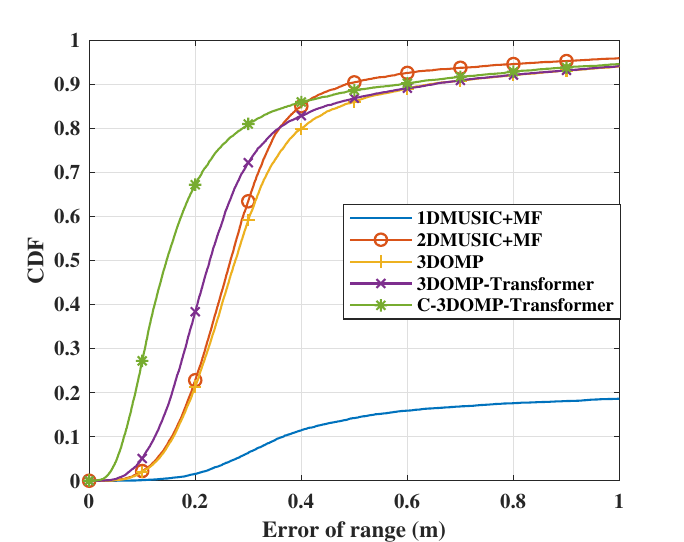}} \\
\subfloat[CDF of the least match]{\includegraphics[width = 3in]{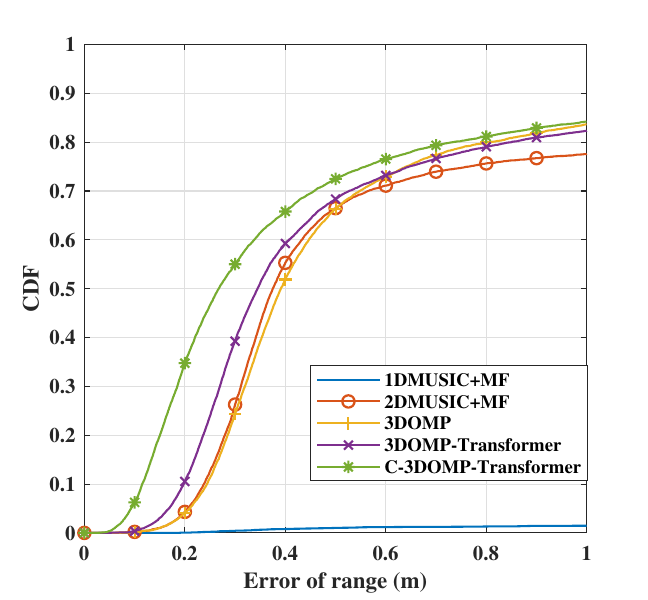}} 
\caption{The CDF of range error from the least match targets to the most match in different methods when $M=5$ and $SNR_{s}=10\, dB$.}
\label{fig:CDF-M=5-SNR10}
\end{figure*}

Fig. \ref{fig:CDF-M=5-SNR10} shows the CDFs of range error from the most match to the least match targets among $M$ targets in different methods when $M=5$ and $SNR_{s}=10\,dB$. From the figure, the proposed 3D-OMP-Transformer and C-3D-OMP-Transformer perform the best in most cases. When evaluating the CDFs of the third and the fourth most match targets, 2D-MUSIC+MF performs better when the error threshold exceeds a certain value. However, the proposed methods still have the smallest average MAE in the two cases. Note that if minimizing the error threshold to reach 90\% CDF is the designing objective, we can realize the objective by slightly changing the loss functions, i.e., Eq. (\ref{equ:29}) and Eq. (\ref{equ:36}). 

\begin{table*}[!ht]
\centering
\caption{The running time comparison of different methods when testing $200$ samples. }
\begin{tabular}{|c|c|c|c|c|c|}
\hline
M & 1D-OMP+MF & 2D-OMP+MF & 3D-OMP & 3D-OMP-Transformer & C-3D-OMP-Transformer\\ \hline
              $1$          &    $8.233\, s$                                                                          &  $24.385\, s$                                                                           &   $6.603\, s$                          &   $7.621\, s$                                                                                  &  $8.729\, s$                                                                                     \\ \hline
                    $5$        &   $9.022\, s$                                                                       &     $24.413\, s$                                                                         &  $7.398\, s$                            &  $14.47\, s$                                                                                  &  $18.08\, s$                                                                                     \\ \hline
\end{tabular}                                           
\label{tab:3}
\end{table*}

Table. \ref{tab:3} shows the running time of different methods for testing $200$ samples with a batch size of two. All the methods are implemented by Pytorch and accelerated by GPUs for a fair comparison. From the table, 3D-OMP is the fastest algorithm. 3D-OMP-Transformer and C-3D-OMP-Transformer have comparable speeds with 3D-OMP when $M=1$. However, due to the introduction of the FFN, the running time of 3D-OMP-Transformer and C-3D-OMP-Transformer is longer than 3D-OMP when $M=5$. 

\subsection{Additional Experiments with Changed Settings}
\begin{table}[!ht]
\centering
\caption{The performance comparison of different methods when $K=64$, $SNR_{s}=10\,dB$, and $M=3$. }
\begin{tabular}{|c|c|c|c|}
\hline
Methods & 3D-OMP & \begin{tabular}[c]{@{}c@{}}3D-OMP\\ -Transformer\end{tabular} & \begin{tabular}[c]{@{}c@{}}C-3D-OMP\\ -Transformer\end{tabular} \\ \hline
$\phi$  &   0.00125     &  0.001055     & 0.000858                      \\ \hline
$r$     &  0.356      &  0.289                  &   0.205                   \\ \hline
$v$     &   0.285     &  0.182                  &   0.177                   \\ \hline
\end{tabular}
\label{tab:4}
\end{table}  
Table \ref{tab:4} shows the performance comparison of different methods when $K=64$, $SNR_{s}=10\,dB$, and $M=3$. Compared with Table. \ref{tab:1}, the sensing accuracy increases when $K$ increases. In this experiment, 3D-OMP-Transformer reduces the angle error, range error, and velocity error of 3D-OMP by $16\%$, $19\%$, and $36\%$, respectively, which is larger than the $6\%$, $13\%$, and $25\%$ when $K=16$ in Table. \ref{tab:1}. C-3D-OMP-Transformer can reduce the angle error, range error, and velocity error of 3D-OMP by $32\%$, $43\%$, and $38\%$, respectively, which is also larger than the $8\%$, $28\%$, and $33\%$ when $K=16$ in Table. \ref{tab:1}. This means 3D-OMP-Transformer and C-3D-Transformer are more likely to get better performance than 3D-OMP when the number of antennas is sufficient. 

\begin{table}[!ht]
\centering
\caption{The performance comparison of different methods when $N_{\phi}=180$, $N_{r}=150$, $N_{v}=30$, $SNR_{s}=10\,dB$, and $M=3$. }
\begin{tabular}{|c|c|c|c|}
\hline
Methods & 3D-OMP & \begin{tabular}[c]{@{}c@{}}3D-OMP\\ -Transformer\end{tabular} & \begin{tabular}[c]{@{}c@{}}C-3D-OMP\\ -Transformer\end{tabular} \\ \hline
$\phi$  &   0.00512     &  0.00424     & 0.00385                      \\ \hline
$r$     &  0.9767      &  0.7984                  &   0.5135                   \\ \hline
$v$     &   0.6183     &  0.3686                  &   0.3097                   \\ \hline
\end{tabular}
\label{tab:5}
\end{table} 
Table \ref{tab:5} shows the performance comparison of different methods when $N_{\phi}=180$, $N_{r}=150$, $N_{v}=30$, $SNR_{s}=10\,dB$, and $M=3$. Compared with Table. \ref{tab:1}, the sensing accuracy decreases when the size of grids and dictionaries decreases. In this experiment, 3D-OMP-Transformer reduces the angle error, range error, and velocity error of 3D-OMP by $17\%$, $19\%$, and $41\%$, respectively, which is larger than the error reduction when $N_{\phi}=360$, $N_{r}=300$, $N_{v}=60$ in Table. \ref{tab:1}. C-3D-OMP-Transformer can reduce the angle error, range error, and velocity error of 3D-OMP by $25\%$, $48\%$, and $50\%$, respectively, which is also larger than Table. \ref{tab:1}. This means 3D-OMP-Transformer and C-3D-Transformer are more likely to get larger gain when the size of grids and dictionaries of 3D-OMP is limited.

\begin{table}[!ht]
\centering
\caption{The performance comparison of different methods when $SNR_{s}=20\,dB$ and $M=3$. }
\begin{tabular}{|c|c|c|c|}
\hline
Methods & 3D-OMP & \begin{tabular}[c]{@{}c@{}}3D-OMP\\ -Transformer\end{tabular} & \begin{tabular}[c]{@{}c@{}}C-3D-OMP\\ -Transformer\end{tabular} \\ \hline
$\phi$  &   0.00284     &  0.00257     & 0.00238                     \\ \hline
$r$     &  0.4342      &  0.3817                  &   0.3019                   \\ \hline
$v$     &   0.3561     &  0.2261                  &   0.1974                   \\ \hline
\end{tabular}
\label{tab:6}
\end{table} 
Table \ref{tab:6} shows the performance comparison of different methods when $SNR_{s}=20\,dB$ and $M=3$. Compared with Table. \ref{tab:1}, the sensing accuracy increases when $SNR_{s}$ increases. Similarly, the performance improvement from 3D-OMP-Transformer and C-3D-OMP-Transformer become larger.

\section{Conclusion}
\label{sec:conclusion}
In this work, we have designed a novel white-box 3D-OMP transformer for ISAC. We find a new mathematical interpretation of transformers from the perspective of 3D-OMP algorithm and provide a promising way to realize an efficient 3D attention mechanism. Besides, we have shown AI-based solutions can perform better than existing MUSIC or OMP-based solutions in multi-target detection task of ISAC. We have also shown that the performance of 3D-OMP can be further improved by introducing learnable parameters.

\bibliographystyle{IEEEbib}
\bibliography{strings,refs}

\end{document}